\documentclass[modern]{aastex63}

\usepackage{bm}

\def\mJB{{\rm mJy~beam^{-1}}}
\def\muJB{{\rm \mu Jy~beam^{-1}}}

\def\kms{{\rm km~s^{-1}}}
\def\Ms{M_{\sun}}

\revised{\today}
\submitjournal{ApJ}

\shorttitle{Disk Identification Methods}
\shortauthors{Aso \& Machida}

\begin{document}
\title{Testing Disk Identification Methods Through Numerical Simulations of Protostellar Evolution}
\correspondingauthor{Yusuke Aso}
\email{yaso@kasi.re.kr}

\author[0000-0002-8238-7709]{Yusuke Aso}
\affil{Korea Astronomy and Space Science Institute (KASI), 776 Daedeokdae-ro, Yuseong-gu, Daejeon 34055, Republic of Korea}

\author[0000-0002-0963-0872]{Masahiro N. Machida}
\affil{Department of Earth and Planetary Sciences, Faculty of Sciences, Kyushu University, Fukuoka 812-8581, Japan}

\begin{abstract}
We test whether the radii of circumstellar disks can be reliably determined in observations applying the results of a numerical simulation.
Firstly, we execute a core collapse simulation which starts from a rotating magnetized spherical core, and continue the calculation until the protostellar mass reaches $0.5~\Ms$.
Then, for each set of simulation data, we calculate the radiative transfer to generate the data cube for the synthetic observation.
The spatial and velocity resolutions of the synthetic observation are $0\farcs 15$ (20 au) and $0.1~\kms$, respectively. 
We define seven different disk radii.
Four radii are estimated from the synthetic observation, using the continuum image, continuum visibility, C$^{18}$O channel map, and C$^{18}$O position velocity (PV) diagram. 
The other three  radii are taken from the simulation and use the disk rotation, infall motion,  and density contrast around the protostar to identify the disk.  
Finally, we compare the disk radii estimated from the systemic observation with those from the simulation.
We find that the disk radius defined using the PV diagram can reliably trace the Keplerian disk when the protostellar mass is larger than $M_*\gtrsim 0.2~\Ms$ independent of the inclination angle to the  line of  sight.
In addition, the PV diagram  provides an accurate estimate of the central stellar  mass through the whole protostellar evolution. 
The simulation also indicates that the circumstellar disk is massive enough to be gravitationally unstable through the evolution.
Such an unstable disk can show either a circular or spiral morphology on a similar timescale.
\end{abstract}


\section{Introduction} \label{sec:intro}
Disks around young stellar objects play an important role in star formation.
They are hosts for planet formation. 
The disks launch protostellar outflows, which are an indicator for identifying protostars. 
In addition, the disk kinematics or Keplerian rotation motion can be used to estimate the mass and evolutionary stage of deeply embedded protostars.
In the classical picture of star formation, the Keplerian rotating disk has an expected size of $\sim100$ au. 
Past observations reported large-sized disks around typical T Tauri stars \citep[see reviewed by][]{wi.ci11} and some protostars \citep[][]{yen17}. 
However, more recent surveys have revealed that T Tauri stars have diversity in their disk size ranging from  tens to hundreds of au \citep{naji18, ansd18}. 
The disk size distribution in the protostellar (or main accretion) phase seems to be different in  different star-forming regions \citep{ciez19, aso19, tobi20}. 
Recent observations have again focused on disk size as a way to comprehensively understand the star formation process. 

Thus, it is important to accurately identify disks in observations in order to investigate their size and mass, which are closely related to the evolutionary stage of the star and planet formation. 
However, there is no established method to identify disks around young stellar objects from observational data. 
It is especially difficult to estimate the disk size around Class 0 and I protostars, because the protostellar systems are embedded in a dense  infalling envelope. 
Currently, several observational studies have proposed different methods to identify disks and to estimate disk size.
For example, researchers often use molecular lines to identify a disk around a single object (or a few objects). 
In this case, using a position velocity (PV) diagram taken from the molecule line data, a power-law index of the rotational velocity is measured as a function of radius \citep[e.g.,][]{lee10, yen13, ohas14, bjer16}. 
The Keplerian rotation velocity is proportional to $r^{-1/2}$, while that of the infalling envelope is proportional to $ r^{-1}$ \citep{yen13,takahashi16}.
In some line observations, the molecular lines were used to inspect the radial difference in the velocity gradient, which determines the boundary between a purely rotating disk and the surrounding envelope \citep[e.g.,][]{aso15, mare20}. 

Survey studies including numerous targets have difficulties detecting molecular line emissions.
In such studies, the full width at half maximum (FWHM) of the continuum emission derived from Gaussian fitting has been adopted for measuring the disk size \citep[e.g.,][]{ciez19, tobi20}. 
However, the consistency of the methods used in these studies has never been examined systematically.
Thus, we cannot fairly compare the disk properties among different observations or different methods. 

A systematic test of the methods for identifying disks is crucial to allow correct estimates of disk size in observations.
However, it is difficult to evaluate which method is most reliable only from observational data because the actual properties of the disks are unknown.
Through theoretical simulations, we can test which methods are most appropriate, by determining the disk physical quantities such as the disk radius and mass without observational limitations. 
In previous observational studies, an analytical model \citep[e.g.,][]{sh.ei17} and a snapshot of a simulation of an observed object \citep[e.g.,][]{taka17} were typically used  to estimate the disk quantities and central stellar mass. 
Recent developments in computers and numerical techniques allow us to fit  observations using many free parameters. 
Such fitting could, however, suffer from degeneracy of the parameters, in addition to the high computational cost. 
It is hence worth establishing a method to clearly characterize the disk properties such as the disk radius and stellar mass through the protostellar evolution.
Parameter surveys and specific case studies used in past  observational studies have not investigated whether such physical quantities can be consistently estimated by a specific method through all stages of protostellar evolution. 
In this study, we test four representative methods for estimating the disk radius in observations, both using protostar formation simulations and synthetic observations. 

The rest of the paper is structured as follows. 
We describe  the settings and results of our simulation and the procedure of radiative transfer and synthetic observation in \S2.
The results and analyses of disk radius are presented in \S3. 
We discuss each method for identifying disks and disk properties in \S 4.
A conclusion is presented in \S5.

\section{Numerical Settings} \label{sec:num}
\subsection{Core Collapse Simulation} \label{sec:sim}
The numerical simulation used in this study is the same as that used in \citet{tomi17}.
This section provides a brief summary of the numerical settings and simulation results.

As the initial state, a spherical cloud core with a critical Bonnor--Ebert density profile is adopted with a central molecular-hydrogen density of $6\times 10^5~{\rm cm}^{-3}$ and an isothermal temperature of 10 K.  
The radius and mass of the cloud core are 0.03 pc and $1.25~\Ms$, respectively.
The cloud core is penetrated by a uniform magnetic field with an initial amplitude of 51 $\mu$G\footnote{
The magnetic field strength was incorrectly wrote in \citet{tomi17} as 36$\mu$G. The mass-to-flux ratio $\mu/\mu_{\rm crit}=3$ is correct in both \citet{tomi17} and this paper.} and rigidly rotates with an initial angular frequency of $2\times 10^{-13}~{\rm s}^{-1}$. 
The initial rotation axis, which is set to be parallel to the $z$-axis, is parallel to the direction of the initial magnetic field.

A nested grid code is used to cover a wide spatial range \citep[for details, see][]{machida04,machida05}. 
Each grid has $64\times 64$ cells  in the $x$ and $y$ directions and  32 cells in the upper half along the $z$-axis.
Note that the lower half mirrors the upper half (i.e., a mirror boundary is imposed on the $z=0$ plane). 
The cell width of the innermost grid is 0.75 au and the box size and cell width of the outermost grid are 1~pc and 0.016~pc, respectively. 
The equations of resistive magnetohydrodynamics (MHD) are solved to derive the density and velocity distributions through the protostellar evolution \citep{machida06}.
Since the barotropic equation of state is used, the gas pressure and temperature are given as functions of  density. 
A fixed outer boundary is set much further out than the initial core radius \citep{machida13}. 
A sink cell with a size of 1~au is imposed  as the inner boundary. 
Parcels of gas that fall into the sink cell are stored as the central protostellar mass, which reaches $0.5~\Ms$ at the end of the simulation.
We calculate the cloud evolution for 0.1 Myr after the cloud collapse begins.

\begin{figure*}[htbp]
\epsscale{1.}
\plotone{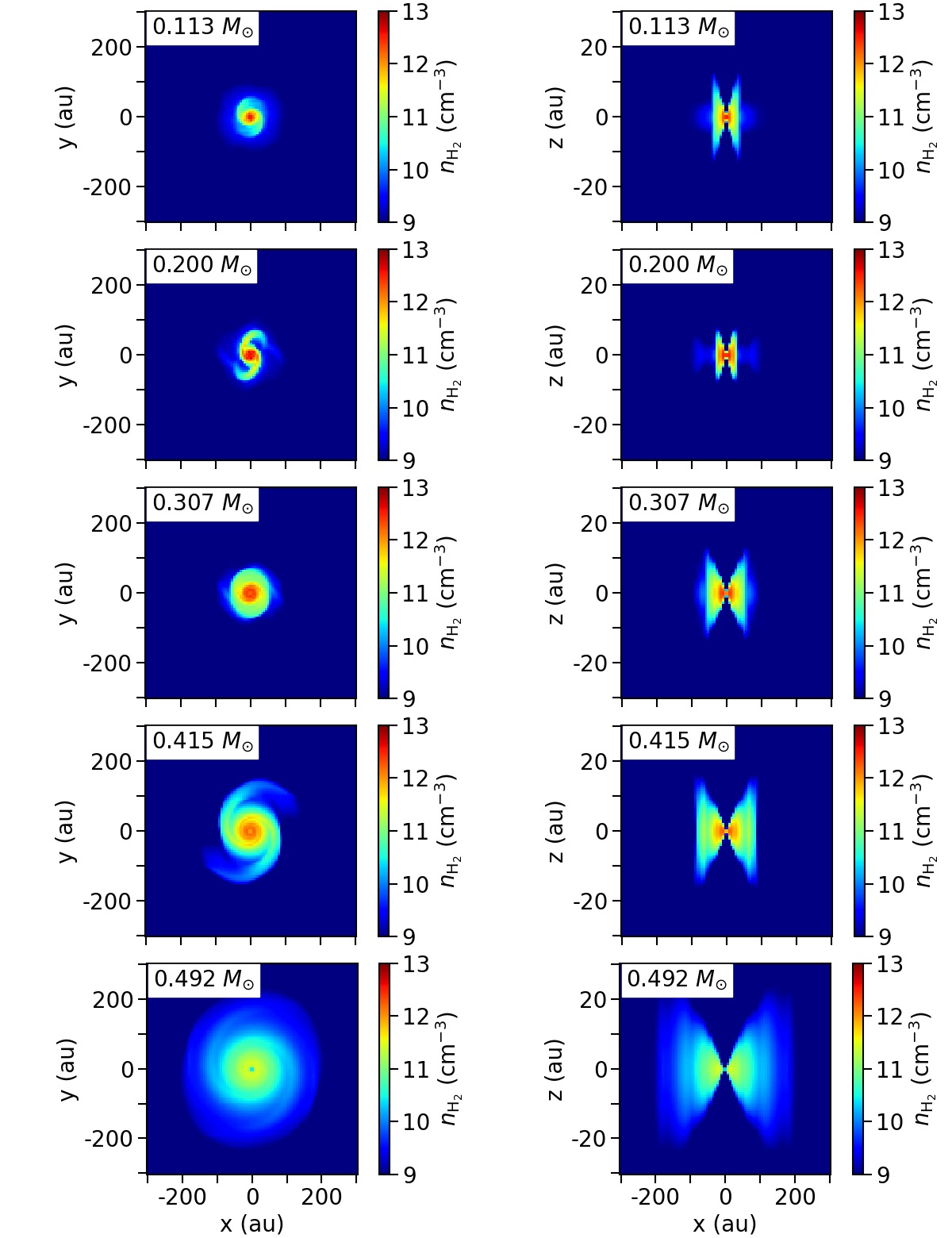}
\caption{
Density distributions at five different epochs in the midplane ({\it left})  and the $xz$ plane including the pole  ({\it right}). 
The central stellar mass at each epoch  is denoted in the top left corner. 
In the right panels, the spatial scale of the $z$-axis is 0.1 times smaller than the other axes.
\label{fig:nT}}
\end{figure*}

Figure \ref{fig:nT} shows the density distribution at five different epochs when the central stellar mass reaches  $M_{\rm *}=0.113$, 0.200, 0.307, 0.415, and $0.493~\Ms$.
Overall, the density decreases as the radius from the center increases, while it also decreases above and below the disk-like structure. 
A diffuse region with a density of $n_{\rm H_2}< 10^9~{\rm cm}^{-3}$ surrounds a dense region ($n_{\rm H_2}\gtrsim  10^9~{\rm cm}^{-3}$) in the $xy$ plane (left panels of Figure~\ref{fig:nT}). The diffuse region likely corresponds to an infalling envelope, while the dense region is likely a disk.
The dense region increases in size until the end of the simulation (left panels of Figure~\ref{fig:nT}), which is clearly seen in the $xz$ plane (right panels of Figure~\ref{fig:nT}). 
The dense region also shows a flared structure with a scale height proportional to the radius.
In addition, the panels labeled $M_{\rm *}=0.200$ and $0.415~\Ms$ show a two-arm spiral structure in the $xy$ plane. 
The other panels ($M_{\rm *}=0.113$, $0.307$, and $0.492~\Ms$) also show such a non-axisymmetric structure.
However, a non-axisymmetric structure at the epochs of $M_{\rm *}=0.113$, $0.307$, and $0.492~\Ms$ is  less clear than at the epochs of $M_{\rm *}=0.200$ and $0.415~\Ms$. 
The density increases in the region of $r\lesssim 200$~au as the system evolves, whereas the peak density decreases after the central mass reaches $M_{\rm *}=0.415~\Ms$ \citep{shu77}.

\clearpage

\subsection{Radiative Transfer and Synthetic Observation} \label{sec:rad}
A core collapse simulation can provide distributions of physical quantities such as density and  velocity, 
which cannot be directly derived from observations. 
After the simulation is  executed,  two further  steps (radiative transfer process and synthetic observation) are required to obtain observable quantities such as intensity and visibility.
We perform a radiative transfer calculation to determine  the intensity along any line of sight for a given distribution of physical quantities in the core collapse simulation. 
We use the open source code RADMC3D \citep{dull12} to calculate the radiative transfer process to determine  the intensity of the C$^{18}$O $J=2-1$ line (219.5603541 GHz) and 1.36 mm (220 GHz) continuum emissions. 
For  the dust continuum emission, we adopt the RADMC3D default dust opacity  $0.002~{\rm cm}^{2}~{\rm g}^{-1}$ at 1.36 mm, which is a typical value in the interstellar medium. 
Since dust grains should evolve in a protostellar system, the dust opacity of the disk and surrounding envelope might be different. 
Thus, it may be necessary to consider the grain growth and dynamical evolution of  the protostellar system consistently.
However, the grain growth and its effects are beyond the scope of the present study
and will be focused on in future studies.
We also set a gas-to-dust mass ratio of 100, which is typical in the interstellar medium. 

The calculation of the C$^{18}$O line emission uses a fractional abundance of C$^{18}$O relative to H$_2$ molecules of $X({\rm C}^{18}{\rm O})=5\times 10^{-7}$ \citep{lacy94, wi.ro94} when the gas temperature is above 20 K.
The abundance is set to decrease by two orders of magnitude when the gas temperature is below 20 K, which reflects the  freeze-out of CO isotopologues onto the surface of dust grains. 
Note that we use the gas temperature adopted in the simulation to keep consistency between the simulation and synthetic observation (\S\ref{sec:sim}).
Both line and continuum opacities are included when calculating the radiative transfer.
In performing the radiative transfer process and synthetic observations, we assume that the location of the simulated object (i.e., protostellar system)  is at the same position on the plane of the sky. For this, we adopt the same distance as the protostar L1527 IRS in the Taurus star forming region: $\alpha (2000) = 04^{\rm h}39^{\rm m}53\fs 9$, $\delta (2000)=+26\arcdeg 03\arcmin 10\arcsec$ \citep{aso17} and $d=140$ pc. 
The inclination angle of the simulated object is  set to be mid-way between the edge-on and face-on orientations, i.e., $i=45\arcdeg$. 
We also adopt two azimuthal angles, $\phi=0\arcdeg$ and $90\arcdeg$, to investigate the dependence on the azimuthal variation of the physical quantities, 
discussed in \S\ref{sec:dis}.
The position angle for the  object in RADMC3D is set to be $0\arcdeg$, indicating that the rotation axis or the disk normal is north in the plane of the sky. 
The derived image has a size of 2048 pixels, where one pixel has a size of $0\farcs 02$. 
The velocity of C$^{18}$O emission ranges  from $-5$ to $5~\kms$ with a velocity resolution of $0.1~\kms$, in which 101 velocity channels are allocated.

After calculating the radiative transfer process,  a continuum model image is constructed as the geometric mean of the images at $\pm 15~\kms$ with respect to the rest frequency of C$^{18}$O ($J=2-1$). 
Similarly,  continuum images in the range $-15$ to $15~\kms$ can be estimated using a power-law function of frequency (velocity). The estimated continuum images are subtracted from the output image cube of RADMC3D to generate the C$^{18}$O image cube.
We execute the continuum subtraction process before performing the synthetic observations to reduce the time for calculating images at velocities that are too high  to be used in our analysis.
This treatment of continuum subtraction does not significantly affect our results shown in this paper as demonstrated in Appendix \ref{sec:app}.

Synthetic observations are made of the visibility (in units of Jy), directly observable in interferometers, from an input model image (in units of Jy$~{\rm pixel}^{-1}$). 
We use the task {\it simobserve} in the Common Astronomy Software Applications (CASA) for this process, adopting a single pointing. 
The input antenna configuration is taken from a public file, alma.cycle7.7.cfg, which corresponds to the antenna configuration C43-7 of ALMA in Cycle 7 and covers a projected-baseline range from 43 to 2980 m. 
The phase reference center is the coordinates of the object. 
The on-source observing time is set to 30 minutes centered at the transit time on January 1 in 2020, which is divided into 10 integrations (each integration time is 3 minutes). 
The synthetic observations do not include artificial noise, though the effect of noise is added in our analyses, as described in \S\ref{sec:ana}. 
The noise inclusion process adopted in this study is different from real observations, while it does not affect our result as demonstrated in Appendix \ref{sec:app}.

The derived visibility is inverse-Fourier transformed and CLEANed to produce an image, in the same way as real interferometric observations, through the task {\it tclean} in CASA.  
The derived image has the same size as the input image derived in the radiative transfer process, which covers the response of the ALMA primary beam down to 20\% at 1.36 mm. 
The CLEAN process is  terminated at thresholds of 0.04 and 2 $\mJB$ for the continuum and line emission cases, respectively. 
Typical weighting is set, namely a Briggs weighting with a robust parameter of 0.5, providing an angular resolution of $0\farcs1498 \times 0\farcs 1148$ (P.A. $=-3.3\arcdeg$). 
For simplicity, the output images are convolved to the final circular beam with an angular resolution of $0\farcs 15$ (20 au) after the CLEAN process.
This spatial resolution is reasonably good for observational studies of protostellar disks. 

In the next section (\S\ref{sec:ana}), we show both cases before and after performing the synthetic observation. 
Compared with the case after performing the synthetic observation with the artificial noise, the case before it can be interpreted as the observation with an extremely high spatial resolution and sensitivity without missing flux.
In other words, the case without the synthetic observation correspond to an ideal  (or unrealistically good) observational setup, while the case with the synthetic observation correspond to a realistic observational setup. 
In this paper, to focus on the main scope (i.e., a systematic test through protostellar evolution), we compare the ideal observational setup with the reasonable setup without adopting other observational setups.


\section{Analysis and Results} \label{sec:ana}
In this section, we define various radii that are supposed to trace the ``disk radius" from the simulation directly or from the model image after  performing the radiative transfer and synthetic observations. 
The radii defined in the simulation show how much the disk radius can vary depending on the physical quantities. 
Comparing the radii defined in the simulation with those derived in the model image, we demonstrate  how precisely the disk radius is defined by observable quantities.   

The radii derived after performing the synthetic observation includes artificial noise.
The noise is generated from a Gaussian random number at each pixel. The random numbers are convolved with a Gaussian beam with a FWHM of $0\farcs 15$ (20 au). 
The noise has a standard deviation of $0.02~\mJB$ for the continuum case and $1~\mJB$ for the C$^{18}$O case. 
They are achievable by ALMA with a reasonable observing time. 
The noise level of $0.02~\mJB$ is equivalent to $0.02\ {\rm mJy}\times \sqrt{N_{\rm bl}}$ for the real part of the visibility for each baseline, where $N_{\rm bl}$ is the number of baselines. 
For $N_{\rm bl}=904$ (see \S\ref{sec:vis}), the noise level is 0.60~mJy. 
Each radius measurement is iterated 100 times with $\phi=0$ and $90\arcdeg$ with different artificial noises to remove the dependency of the disk radius on the azimuthal direction $\phi$.

\subsection{Radii Estimated from Simulation} \label{sec:true}
The term ``disk" commonly means a rotationally supported disk or a Keplerian disk. 
Hence, the  disk radius should correspond to  the radius within which the material  is supported by rotation against  gravity.
The balance between  rotation (or centrifugal force) and gravity also indicates that the material in a disk has no radial velocity component, i.e., no infall (or expansion) velocity. 
In contrast, the material surrounding the disk falls onto the disk surface, causing  a density jump (or density contrast) that appears near the disk outer edge, in which the density is high inside the density jump and low outside. 
The rotational support, infall motion, and density contrast  can each provide a different definition of the disk in terms of the radius.
In the following, we define three disk radii $R_{\rm rot}$, $R_{\rm inf}$, and $R_{\rm den}$ which are determined from the core collapse  simulation. 

\begin{figure*}[htbp]
\epsscale{1.17}
\plotone{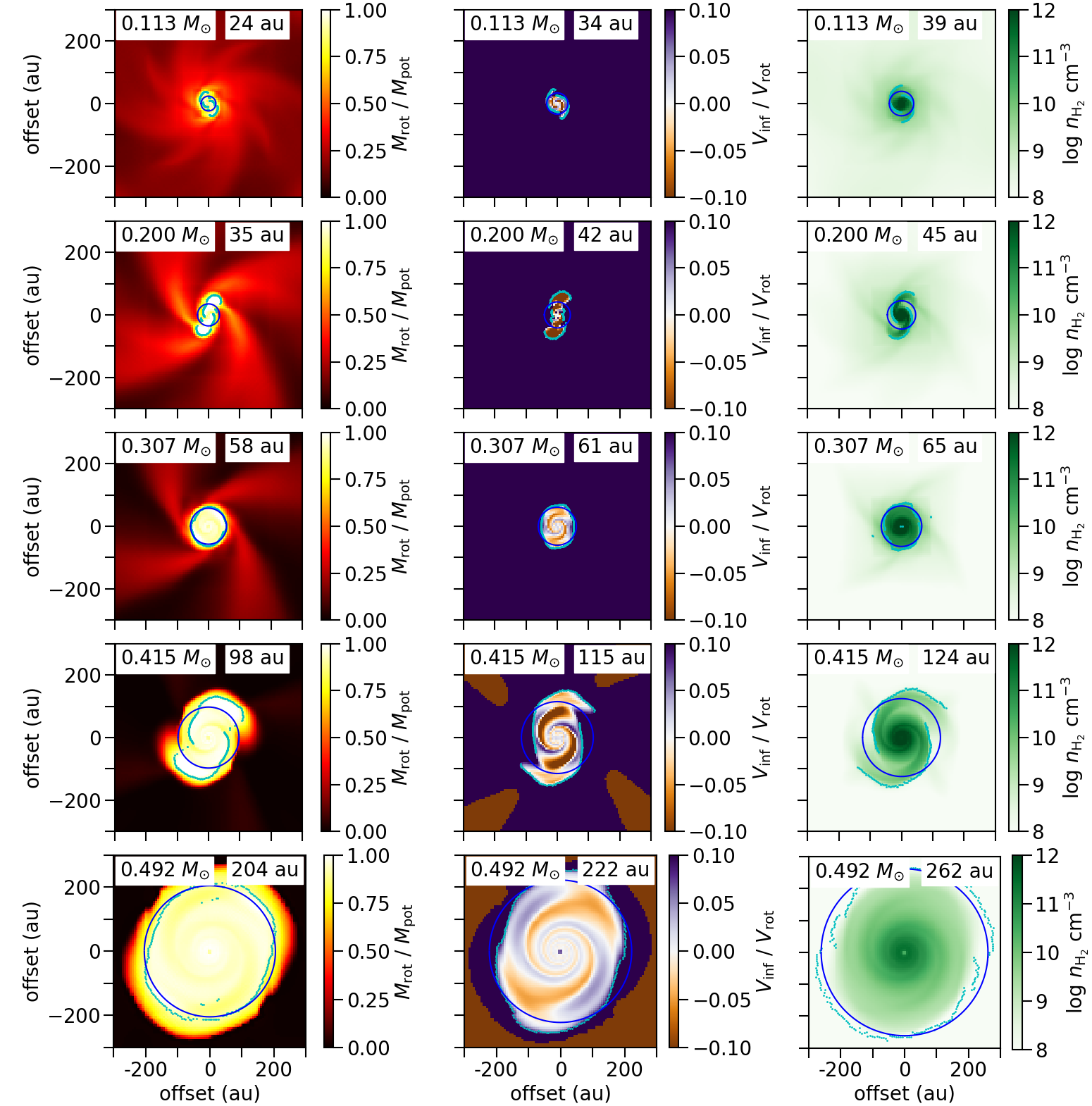}
\caption{
Three radii, $R_{\rm rot}$, $R_{\rm inf}$, and $R_{\rm den}$, defined from physical quantities on the midplane in the simulation. 
The epochs are the same as in Figure \ref{fig:nT}. 
The central stellar mass is denoted in the top left corner of each panel. 
In each panel, the light blue points represent the radius defined in each azimuthal direction, while the blue circle indicates the azimuthally averaged radius. 
{\it Left}: $R_{\rm rot}$, the outermost radius within which  the centrifugal force is stronger than 90\% of the gravitational force, $M_{\rm rot} \ge 0.9M_{\rm pot}$, where $M_{\rm rot}$ and  $M_{\rm pot}$ are defined as $M_{\rm rot} \equiv rv_{\rm rot} ^2 / G$ and $M_{\rm pot}\equiv (d\phi / dr) r^2/G$, respectively.
The color map indicates the ratio $M_{\rm rot}/M_{\rm pot}$.
{\it Middle}: $R_{\rm inf}$, the outermost radius within which  the infall velocity is smaller than 10\% of the rotational velocity, $v_{\rm inf} \le 0.1v_{\rm rot}$. 
The color map indicates  the ratio of  the infall and rotation velocities. 
{\it Right}: $R_{\rm den}$, the radius at which  $-d\ln \rho/d\ln r$ has a (local) peak where the density changes most significantly. 
The color map indicates the density. 
The measured radius is denoted in the top right corner in each panel.
\label{fig:true}}
\end{figure*}

Firstly, we describe the disk radius $R_{\rm rot}$ determined by  the rotation velocity or the balance between the gravitational and centrifugal forces. 
To derive the radius $R_{\rm rot}$, we estimate the $r$-component of the forces in the midplane. 
The gravitational force is calculated from the gravitational potential as  
\begin{eqnarray}
F_{\rm grav}&=&-\rho (\bm{r}) \frac{d\phi}{dr} \nonumber \\
&=&\rho (\bm{r}) \frac{d}{dr}\int \frac{G(\rho (\bm{r}')+M_* \delta(\bm{r}'))}{|\bm{r}'-\bm{r}|}dV' \nonumber \\
&\equiv&-\rho(\bm{r})\frac{GM_{\rm pot}}{r^2},
\end{eqnarray}
where $F_{\rm grav}, \rho, \phi, \bm{r}$, and $r$ are the $r$-component of the gravitational force, density, gravitational potential, position vector, and radius, respectively.  $\delta(\bm{r}')$ is the Dirac delta function.
The mass $M_{\rm pot}$ is defined from the gravitational force, in which the central stellar mass $M_*$ is included. The centrifugal force is calculated from the rotational velocity as 
\begin{eqnarray}
F_{\rm cent}=\rho(\bm{r})\frac{v_{\rm rot}^2}{r} \equiv \rho(\bm{r})\frac{GM_{\rm rot}}{r^2},
\end{eqnarray}
where $F_{\rm cent}$ and $v_{\rm rot}$ are the $r$-components of the centrifugal force and rotational velocity, respectively, and $M_{\rm rot}$ is a mass defined from the centrifugal force. 
We determine the radius of the rotationally supported disk as the outermost point from the protostar for which $M_{\rm rot} \ge 0.9M_{\rm pot}$.
Because this radius is different in each azimuthal direction in the midplane, we define the azimuthally averaged radius as the disk radius, $R_{\rm rot}$. 
The left column of Figure \ref{fig:true} shows $R_{\rm rot}$ overlaid on the distribution of the ratio $M_{\rm rot}/M_{\rm pot}$ in the midplane.

Next, we describe  $R_{\rm inf}$, which is determined by  the infall and rotation velocities. 
It is considered that the infall velocity, $v_{\rm inf}$, is sufficiently small within the disk. 
Thus, we determine the radius as the outermost point from the protostar within which the infall velocity is below 10\% of the rotation velocity, i.e., $v_{\rm inf} \le 0.1v_{\rm rot}$.
Because this radius also has an azimuthal variation, we define $R_{\rm inf}$ as the azimuthally averaged value in the midplane. 
The middle column of Figure \ref{fig:true} shows $R_{\rm inf}$ overlaid on the distribution of the ratio $v_{\rm inf} / V_{\rm rot}$ in the midplane. 
In the panels, the positive sign of the velocities $V_{\rm inf}$ and $V_{\rm rot}$ means the infall motion and counterclockwise rotation, respectively.

Finally, we describe $R_{\rm den}$, determined by the density contrast in the midplane.
Because it is expected that the circumstellar disk has a density profile described as a power-law function of $r$ overall, we define the radius as the peak of the power-law index $-d\ln \rho /d\ln r$ in each direction. 
Then, the radius is azimuthally averaged and defined as  $R_{\rm den}$.
The right column of Figure \ref{fig:true} shows $R_{\rm den}$ overlaid on the distribution of the molecular-hydrogen number density in the midplane.

These radii are denoted in the top right corner of each panel in Figure \ref{fig:true}. 
The figure indicates that  $R_{\rm den}$ is largest and  $R_{\rm inf}$ is larger than $R_{\rm rot}$ at any epoch (i.e., $R_{\rm den}>R_{\rm inf}> R_{\rm rot}$).
When a clear spiral appears  (e.g., for $M_{\rm *}=0.200$ and $0.415~\Ms$), the rotation motion dominates the infall motion within the spiral.
The downstream or the outer part of the spiral is infalling ($v_r < 0$),  while the upstream or the inner part is expanding ($v_r >0$). 
The outer part at the epochs of $M_{*}=0.415$ and $0.492~\Ms$ has ($V_{\rm inf}/V_{\rm rot})<0$ due to clockwise rotation $V_{\rm rot}<0$, rather than due to expanding motion $V_{\rm inf} < 0$.

\newpage

\subsection{Radii from Continuum Images} \label{sec:img}
The continuum emission is considered to better trace disks around YSOs, because the disk has a  higher density than the surrounding medium. 
Thus, the apparent size of the continuum emission, such as FWHM, is often referred to as the disk size. 
Hence, as has been done in many observations,  we fit the continuum images with a single 2D Gaussian function in the synthetic observation. 
Since the size along the major axis is supposed to trace the disk size regardless of the inclination angle, we define a radius $R_{\rm img}$ as the HWHM along the major axis that is derived from the Gaussian fitting. 
In  the synthetic observation, the model Gaussian image is convolved with the (Gaussian) beam for the fitting, so that a deconvolved HWHM is derived as $R_{\rm img}$. 
All the emissions within $\pm 300$ au are used for the fitting. 
The free parameters in the fitting are the amplitude, central coordinate along the declination axis ($y$-axis), major-HWHM, and minor-HWHM, whereas the central coordinate along the right-ascension axis ($x$-axis) and the position angle are fixed, so as to focus on the measurement of the size of the major-axis.

\begin{figure*}[htbp]
\epsscale{1}
\plotone{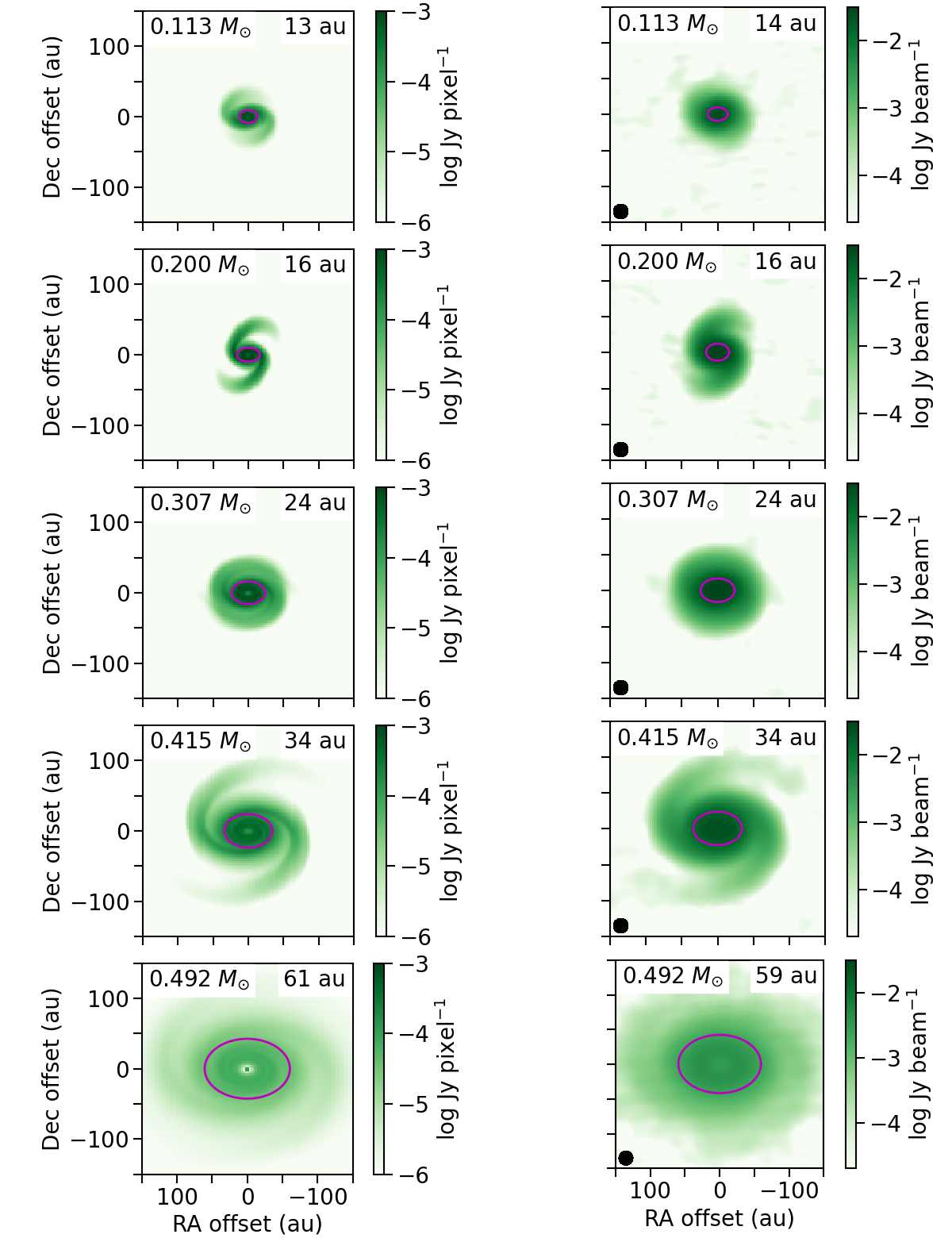}
\caption{
Radius defined from continuum emission in the image domain.
The epochs are the same as in Figure \ref{fig:true}.
$\phi=0\arcdeg$ is adopted. 
The central stellar mass  is denoted in the top left corner of each panel. 
{\it Left}: Continuum emission without  the synthetic observation. 
{\it Right}: Continuum emission with the synthetic observation. 
The purple ellipses represent  radii derived from two-dimensional Gaussian fitting. 
The right column includes artificial Gaussian noise with $\sigma =0.02~\mJB$; its color bar starts at $0.02\ \mJB$. 
The filled black ellipse in the left column denotes the beam size of the synthetic observation, $0\farcs 15$ (20~au). 
The radius $R_{\rm img}$ is denoted in the top right corner in each panel.
\label{fig:img}}
\end{figure*}

Figure \ref{fig:img} shows the Gaussian fitting of the images before and after performing the synthetic observation for $\phi=0\arcdeg$, in which the best-fit $y$-axis position, major-HWHM, and minor-HWHM are shown as the $y$-axis position, semi-major axis, and semi-minor axis of the purple ellipses. 
When the central stellar mass is $M_* = 0.200$ and $0.415~\Ms$,  the continuum emission also shows a clear spiral, as seen in the density distribution with and without the synthetic observation. 
The intensity decreases during the epoch from $M_* = 0.415$ to $0.492~\Ms$, as expected from the density decrease in Figure~\ref{fig:nT}.
The derived radii seem to be  much smaller than the emitting area in Figure \ref{fig:img}. 
Actually, the radii $R_{\rm img}$ are smaller than the radii defined in the simulation $R_{\rm rot}$, $R_{\rm inf}$, and $R_{\rm den}$ by a factor of 2--3. 
Meanwhile, this radius is almost the same between the cases before  and after performing the synthetic observation. 
The purple ellipses in Figure \ref{fig:img} appear to cover only the regions inside the spirals when the spiral is  clear. 

\clearpage

\subsection{Radii from Continuum Visibility} \label{sec:vis}
\begin{figure*}[b!]
\epsscale{1.175}
\plotone{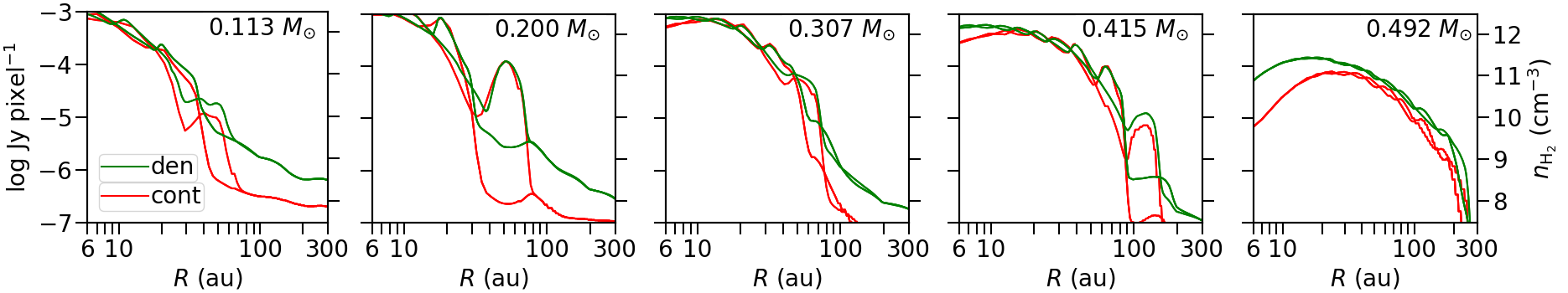}
\caption{
Midplane density with two different azimuthal angles 0 and 90$\arcdeg$ (green lines; right axis) and continuum intensity along the major axis adopting  $\phi=0$ and $90\arcdeg$  before performing the synthetic observation (red lines; left axis) against the radius $R$.
\label{fig:dencont}}
\end{figure*}

The structure of continuum emission is often discussed in the Fourier domain or the $uv$ domain in interferometric observations in order to investigate the data before non-linear image processing such as CLEAN has been performed. 
The Gaussian fitting in the $uv$ domain produces almost the same results as those in the image domain in our settings with  $\sim 1000$ baselines.
Thus, we inspect another method of size measurement, investigating null points in the amplitude or the real part of the visibility.
Note that  the imaginary part of the visibility is zero when the structure is point-symmetric. 
Such null points in the visibility distribution indicate that the intensity distribution has a plateau-like structure with a sharp edge, as discussed in \citet{aso18}. 
Figure \ref{fig:dencont} indicates that the continuum intensity suddenly drops at a radius where the midplane density also rapidly decreases. 
The purpose of this analysis is to identify the radius at which the density contrast becomes very high (i.e., the density jump) using the continuum visibility.
The null point in the visibility is related to the radius of the density jump.
The simplest example is the Fourier transform of a uniform disk, which is proportional to $J_1(1.22\pi\beta / \beta_1)/\beta$ as a function of the $uv$-distance $\beta$ and has the first null point at $\beta = \beta_1$. 
Using the null point, we define $R_{\rm vis}$ as the radius at which a uniform disk produces the first null point, i.e., $R_{\rm vis}=0.6098/\beta_1$, where  $R_{\rm vis}$ is in the units of radians, and $\beta_1$ is the ratio between the baseline length and the observed wavelength.

\begin{figure*}[htbp]
\epsscale{1}
\plotone{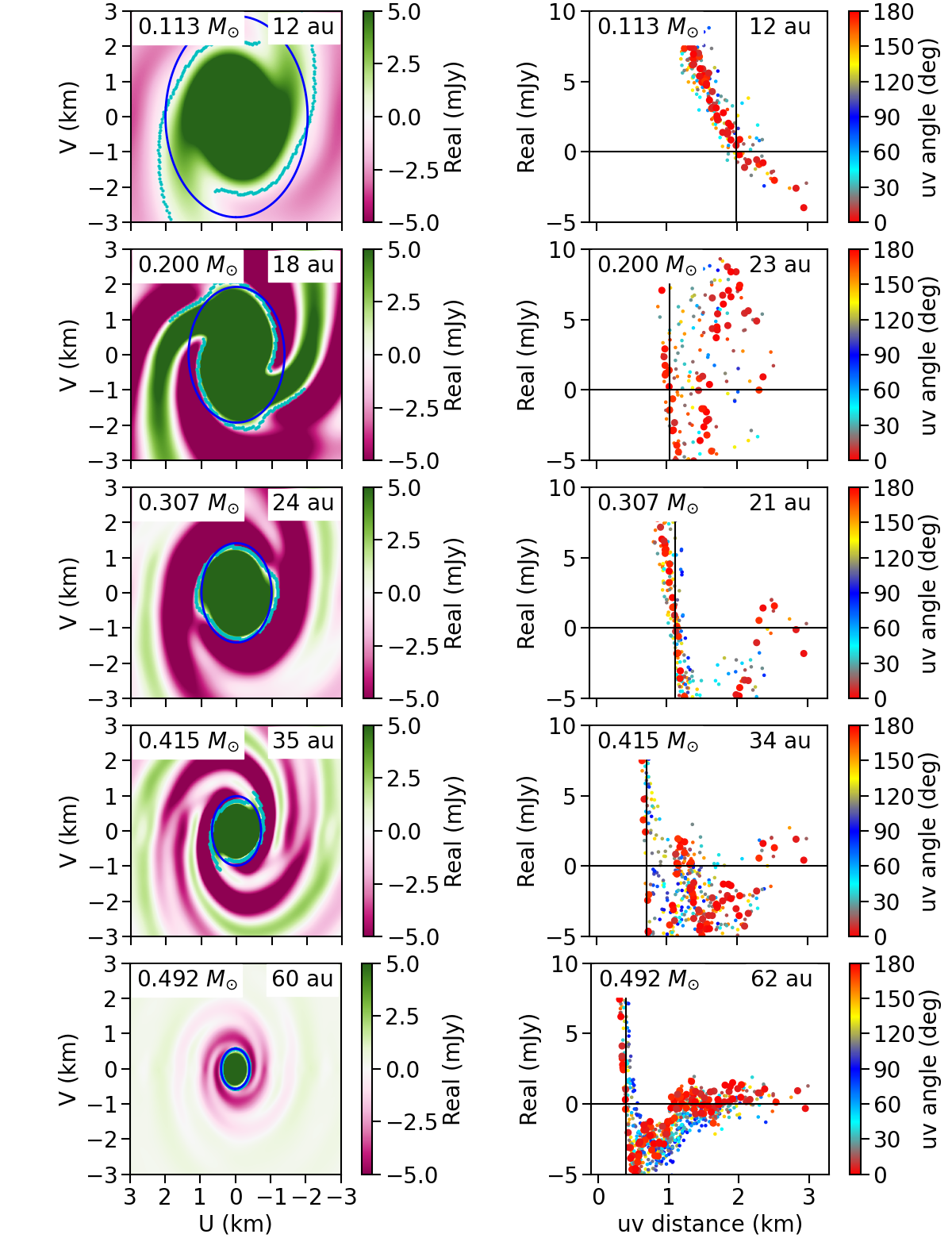}
\caption{ 
Radius defined using the continuum visibility in the $uv$ domain, $R_{\rm vis}$. 
The epochs are the same as in Figure \ref{fig:true}.
$\phi=0\arcdeg$ is adopted. 
The central stellar mass at each snapshot is denoted in the top left corner. 
{\it Left}: Real part of continuum visibility before performing the synthetic observation. 
The light blue points correspond to the innermost null points in different azimuthal directions, while the blue circle shows their average. 
{\it Right}: Real part of the visibility sampled with an antenna configuration of ALMA including an artificial Gaussian noise with $\sigma = 0.60~{\rm mJy}$. 
The vertical line shows the innermost null point along the major axis ($u$-axis$ \pm 10 \arcdeg$). 
The color of each point indicates the azimuthal variation  from the major axis.
The major-axis range is highlighted with a larger point size.
\label{fig:vis}}
\end{figure*}

The left column of Figure \ref{fig:vis} shows the Fourier transform for the model image without  the synthetic observation, which corresponds to  the left column of Figure \ref{fig:img}. 
Clearly, the real part of the visibility  has both positive and negative signs, and thus there exist null points. 
The first null points delineate an elliptical shape in this case, in which the visibility is completely derived on the $uv$ plane. 
Thus, the $uv$-distances of the first null points are derived in all the azimuthal directions and can be azimuthally averaged. 
The averaged $uv$-distance has an ellipse-like shape in the left column of Figure \ref{fig:vis} rather than a circle  because of the inclination angle ($i = 45\arcdeg$) effect. 
The left column of Figure \ref{fig:vis} also shows the null point and a spiral in the opposite direction to the original images (see the left column of Figure \ref{fig:img}). 
This is because the more compact structure at the center in the image domain corresponds to a more extended structure located at the center in the $uv$ domain, which transforms a trailing spiral in the image domain to a leading spiral in the $uv$ domain.

As for real observations, the synthetic observation samples the visibility within the  limited $uv$ coverage determined  by the antenna configuration. 
The CASA tasks {\it simobserve} and {\it listvis} are used for this process.
The integration time and the scan time are the same, 30 minutes, which means that the sampled visibility data are averaged over the whole integration time. Thus, each data point corresponds to each pair of antennas. 
The total number of antennae of 43 provides 903 baselines.
Such a large number of baselines of ALMA enable us to generate the visibility distribution in the 2D $uv$ domain, i.e., the radial and azimuthal directions, even after  time averaging. 

The right column of Figure \ref{fig:vis} shows the sampled visibility data. 
The null points can be clearly seen even after the synthetic observation is performed.
Supposing that the inclination angle may not be known in real observations, we use only the data points near the major axis ($u$-axis$\pm 10\arcdeg$) to determine the $uv$-distance of the first null point. 
The data points in this angle range are highlighted with a larger point size in the right column of Figure \ref{fig:vis}. 
The derived $uv$-distance is indicated with a vertical line in each panel.  
The radius $R_{\rm vis}$ is overall closer to $R_{\rm img}$ than to the radii estimated from the simulation ($R_{\rm den}$, $R_{\rm inf}$,  and  $R_{\rm rot}$). 
In addition, $R_{\rm vis}$ at the epoch of $M_* = 0.200~\Ms$ is larger than that at the next epoch of $M_* = 0.307~\Ms$, whereas this is not the case before the synthetic observation has been performed (left column in Figure \ref{fig:vis}). 
The strong spiral at this epoch shifts the first null point inward along the major axis, compared to those in other directions. 
This causes $R_{\rm vis}$ to be underestimated in the measurement after performing the synthetic observation.

\clearpage

\subsection{Radii from Channel Maps} \label{sec:chan}
The Keplerian disk has no infall (or negative radial) velocity because the centrifugal and gravitational forces are balanced. 
If the disk is inclined, pure Keplerian rotation can provide a velocity gradient along the major axis of the disk.
In contrast, the infall motion in an inclined disk-like envelope provides a velocity gradient along the minor axis.
The combination of rotation and infall motions  provides a velocity gradient. 
Therefore, the velocity gradient along the major axis has been used to identify circumstellar disks in the literature \citep[e.g.,][]{mare20}.

\begin{figure}[htbp]
\epsscale{0.8}
\plotone{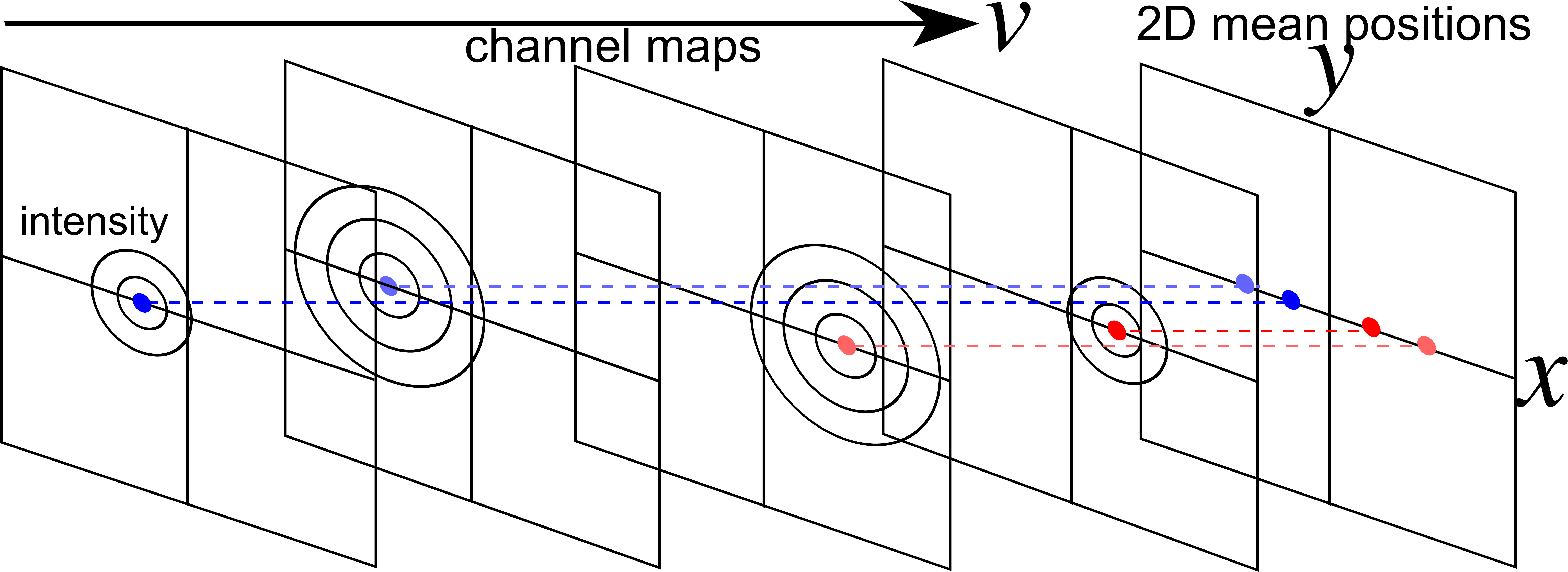}
\caption{
Schematic view of 2D mean positions derived from channel maps. 
The left four panels are channel maps with velocity $v$, in which the contours indicate the 2D intensity distribution and the blue and red points correspond to the 2D mean positions of blue- and red-shifted velocities on the $xy$ plane, respectively.
The rightmost panel is the mean position map.
\label{fig:2dm}}
\end{figure}

\begin{figure}[htbp]
\epsscale{0.7}
\plotone{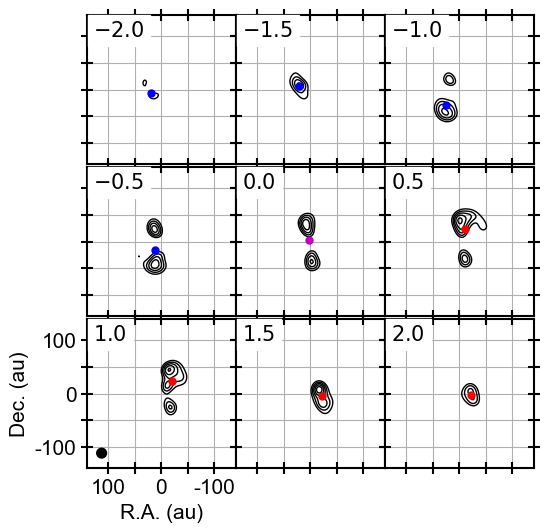}
\caption{
Channel maps at the epoch of $M_*=0.200~\Ms$ after performing the synthetic observation.
The contours are in steps of $2\sigma$ from $5\sigma$, where $1\sigma$ corresponds to $1~\mJB$. 
The line-of-sight velocity is denoted in the upper left corner in each panel. The 2D mean positions are indicated by the blue, purple, and red points.
Although the data cube has a velocity resolution of $0.1~\kms$, only six channels are picked up here.
\label{fig:simchan}}
\end{figure}

The velocity-gradient direction can be determined using a representative position in 2D space in each channel, such as the intensity-peak position or the intensity-weighted mean position. 
In this study, we adopt the 2D mean position described as 
\begin{equation}
 (x_m, y_m) \equiv  \frac{\int \int (x, y) I(x, y) dxdy}{\int \int I(x,y)dxdy},
\end{equation}
 where $(x_m, y_m)$, $(x, y)$, and $I$ are the mean position, spatial coordinates (R.A., Dec.), and intensity as a function of the coordinates, respectively. 
The derivation  of the 2D mean positions is schematically plotted in Figure \ref{fig:2dm}. 
In this study,  the emission is integrated over  $\pm 300$ au$\times \pm 300$ au to determine the 2D mean position.
After the synthetic observation has been performed,  only emissions above $5~\mJB$, which corresponds to $5\sigma$ of the artificial Gaussian noise, are used. Figure \ref{fig:simchan} shows an example of channel maps used in our analysis.

\begin{figure*}[htbp]
\epsscale{1.16}
\plotone{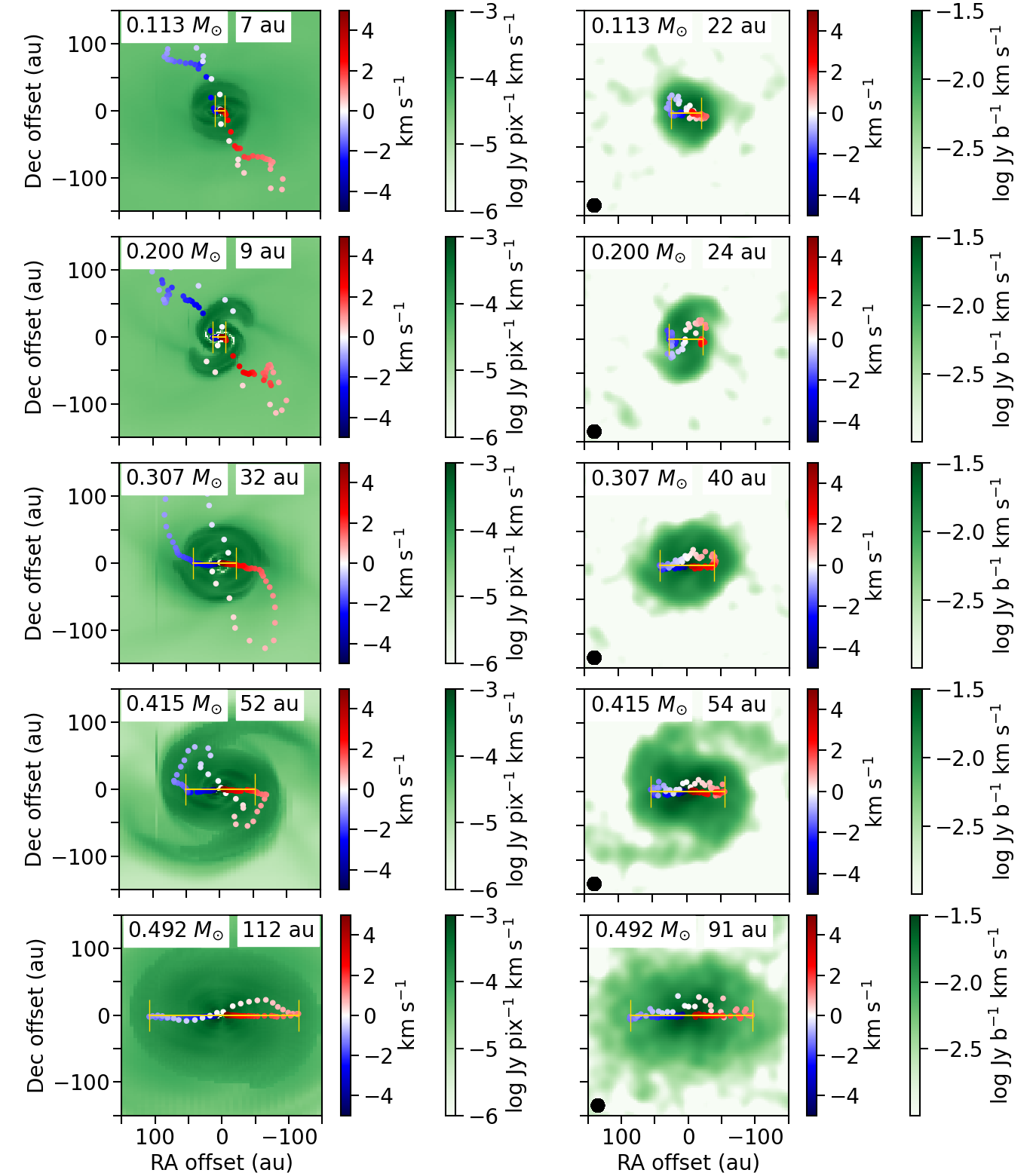}
\caption{
The radius defined in the line channel map before (left panels) and after (right panels) performing the synthetic observation. 
The epochs are the same as in Figure \ref{fig:true}.
$\phi =0\arcdeg$ is adopted. 
The central stellar mass at each epoch  is denoted in the top left corner. 
The intensity-weighted mean positions are plotted with color denoting the velocity. 
In each panel, the yellow bar corresponds to the outermost radius within which the mean positions are along the major axis (i.e., RA-axis). 
The background green color indicates the integrated intensity. 
In the right column,
the filled ellipse at the bottom left corner is  the beam size of the synthetic observation, $0\farcs 15\ (20$ au).
\label{fig:chan}}
\end{figure*}

Figure \ref{fig:chan} shows the derived mean positions, with color denoting the line-of-sight (LOS) velocity overlaid on integrated intensity maps 
before (left panels) and after (right panels) performing the synthetic observation. 
For high velocities, the positions plotted in the figure indicate that the velocity gradient is along the major axis.
In particular, the velocity gradient is clear at the epochs of $M_*=0.307$, 0.415, and 0.492~$\Ms$ in both the left and right panels. 

After performing the synthetic observation (right panels),  the velocity gradient of the low-velocity components is roughly along the major axis at the epochs of $M_*=0.307$, 0.415, and $0.492~\Ms$.
These low-velocity components, however, shift outward as the velocity increases, unlike the high-velocity components. 
This is because, at low velocities, the region with the same LOS velocity (i.e., an iso-velocity contour) is extended beyond the disk. Outer radii beyond the disk has fainter emission and thus less contribute to the mean position than radii in the disk.
In contrast, at high velocities, the iso-velocity contours lie only within the disk. 
The velocity for which  the mean position reaches the outermost radius includes emissions from the outer edge of the disk. 
Meanwhile, emissions also arise from smaller radii than the disk radius, which shifts the mean position inward. 
\citet{aso15} analytically evaluated this shift of the mean position due to emissions at inner radii and found a shifting factor of 0.760.
For this reason, we define the radius $R_{\rm ch}$ as the outermost mean position along the major axis adjusted by the correction factor 1/0.760. 

In the analysis, the mean position is considered to be along the major axis when it is within 1 pixel, $\pm 3$~au, from the major axis. 
The derived radii are shown with yellow bars in Figure \ref{fig:chan}. 
In the early phases, $R_{\rm ch}$ is much smaller before performing the synthetic observation than after it. 
In the later (or evolved) phases, there is no significant difference in $R_{\rm ch}$ before and after performing the synthetic observation. 
In addition, in the later phases, $R_{\rm ch}$ is closer to the radii estimated from the simulation than to the radii estimated from the continuum image and visibility (for details, see \S\ref{sec:dis}).

Except for the epochs of $M_*=0.415$ and $0.492~\Ms$, the mean positions before performing the synthetic observation extend more than those after it,  because the resolving-out effect and the $5\sigma$ cutoff exclude extended components from the measurement in the synthetic observation. 
The infall$+$rotation motion of the envelope in our simulation can create a velocity gradient from the top left to the bottom right, which can be clearly seen, for example, at the epoch of $M_* = 0.113~\Ms$ both before and after performing the synthetic observation. 
The velocity gradient can be confirmed in the perpendicular direction at the epoch of $M_*=0.200~\Ms$ at low velocities in the synthetic observation.
This velocity gradient is not due to the expanding motion but due to the existence of the strong spiral, which can also be seen in the integrated intensity map. 
An analysis using the channel map provides the outermost mean position as being along the major axis.  
The absolute value of the $x$-component of the outermost position is defined as $R_{\rm ch}$. 
Denoting the LOS velocity of the mean position as $v_{\rm ch}$, $R_{\rm ch}$ and $v_{\rm ch}$ provide an estimate of the central stellar mass, $R_{\rm ch}v_{\rm ch}^2 / G / \sin ^2 i$, where $i$ is the inclination angle.
On the other hand, the mean position detected at the highest velocity channel, $(r_h, v_h)$, can provide another estimate of the central stellar mass, $r_{h}v_{h}^2 / G / \sin ^2 i$. 
A reasonable value of the central stellar mass should be between these innermost and outermost estimates. 
Hence, we define the geometric mean of these estimates as $M_{\rm *ch}$ and compare it with the actual stellar mass in \S\ref{sec:mass}.

\clearpage

\subsection{Radii from PV Diagrams} \label{sec:pv} 
The PV diagram enables us to investigate the relation between the LOS velocity and a specific spatial direction. 
Particularly, the PV diagram along the major axis of a (Keplerian) disk gives  the relation between the rotational velocity of the disk and the distance from the central star.
A velocity profile that follows the radius to the power of $-1/2$ (i.e., the Keplerian rotation law) is  direct evidence of a Keplerian disk. 
Such an analysis has been developed and demonstrated in observational studies.
However, the method used to estimate the velocity profile differs among the literature. Examples  include peak positions \citep{lee10,tobi12,yen13,aso15,aso17,sai20} and peak velocities \citep{yen13,lee18,sai20} in a PV diagram, outer edges of emission in a PV diagram \citep{alve17}, a combination of peaks and edges \citep{mare20}, and peak positions in channel maps \citep{hars14,bjer16}.
In other words, the method to estimate the (Keplerian) disk from the PV diagram has not been established. 
In this subsection, we verify the method for accurately obtaining the radial profile of the rotational velocity directly from observations using the PV diagrams derived from our model cube in the C$^{18}$O line.

\begin{figure}[htbp]
\epsscale{0.35}
\plotone{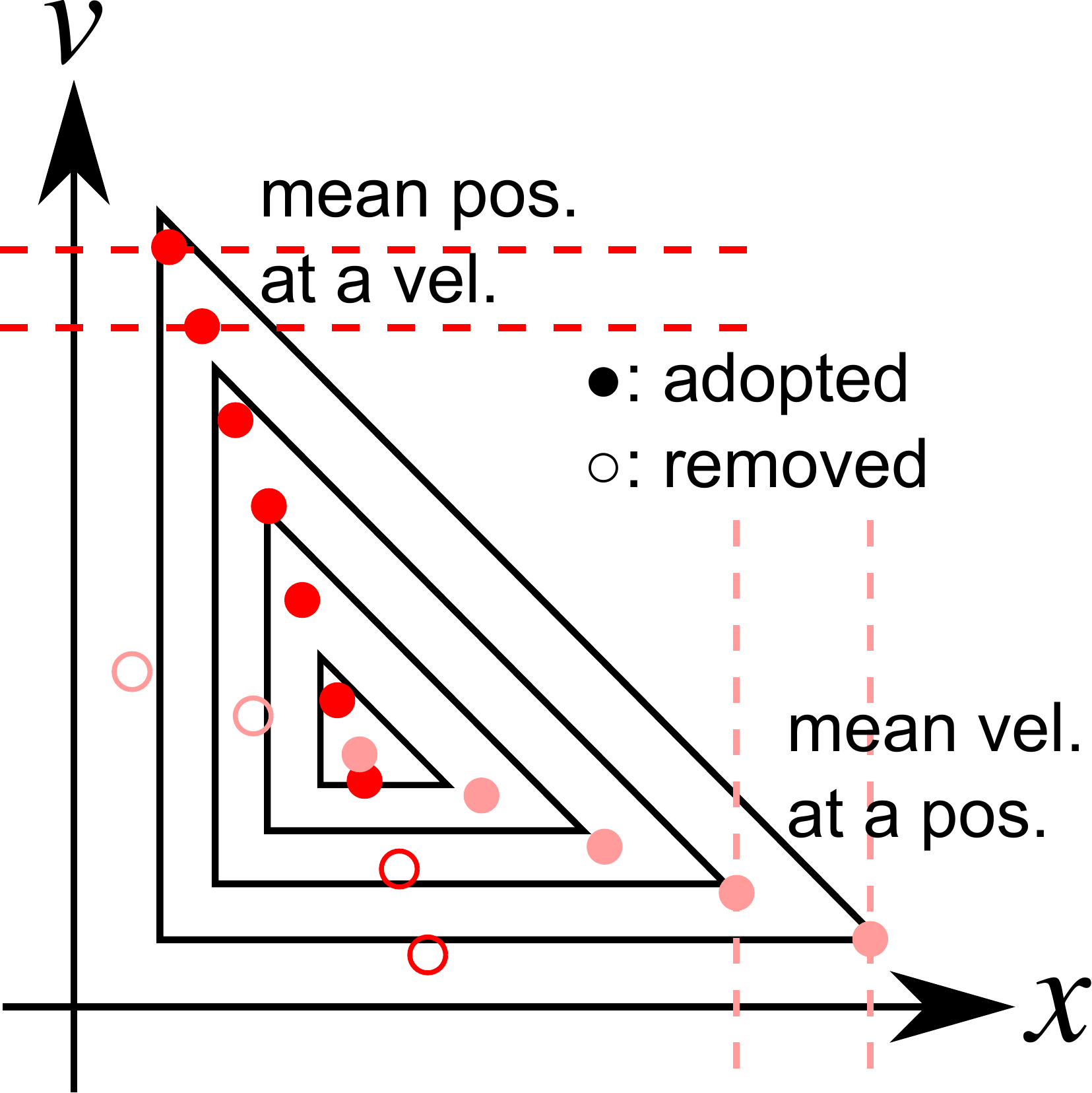}
\caption{
Schematic view of the method to derive 1D mean positions and velocities from a PV diagram. 
The triangles are schematic contours of the PV diagram (first quadrant). 
The red points are the 1D mean position at each velocity channel. 
The pink points are the 1D mean velocity at each position. 
A mean position (red point) is removed when its velocity is lower than the closest counterpart of mean velocity, while a mean velocity (pink point) is removed when it is closer to the center than the closest counterpart of mean position. 
\label{fig:pvm}}
\end{figure}

To investigate the properties of  rotation, such as spin-up (i.e., the rotation velocity increases as the radus decreases), we need to derive a series of representative points tracing the rotation velocity from the PV diagram. 
As described above, there are several different methods  to set the representative points. 
For this purpose, we define  the intensity-weighted 1D mean position at each velocity channel, 
\begin{equation}
    x_m(v)\equiv \frac{\int xI(x, v)dv}{\int I(x, v)dv},
\end{equation}
and the mean velocity at each position,
\begin{equation}
 v_m(x)\equiv \frac{\int vI(x, v)dx}{\int I(x, v)dx},
\end{equation}
where  $x$, $v$, and $I(x, v)$ are the positional coordinate, velocity coordinate and intensity distribution of the PV diagram, respectively.
This method of using the mean points has been analytically discussed in \citet{aso15} and \citet{sai20}. 
The mean velocity is derived in half-beam steps for the Nyquist sampling. 
We only use emissions above $5~\mJB$ ($5\sigma$)  after performing the synthetic observation  to calculate the mean position. 
The mean position preferentially traces higher velocities (inner positions) because of the spin-up feature.
On the other hand, the mean velocity preferentially traces the outer positions (lower velocities). 
To adequately estimate the rotation velocity, we consider both cases, higher velocities/inner positions and lower velocities/outer positions, and determine the representative points using the closest pair of the mean position and mean velocity. 
As shown in Figure \ref{fig:pvm}, either the mean position or the mean velocity is adopted as a representative point in this method. 
The mean position is removed when the velocity of the mean position is lower than that of the closest pair of the mean velocity, while the mean velocity is removed when the position of the mean velocity is closer to the center than that of the closest pair of the mean position (Figure~\ref{fig:pvm}). 

\begin{figure*}[htbp]
\epsscale{1.17}
\plotone{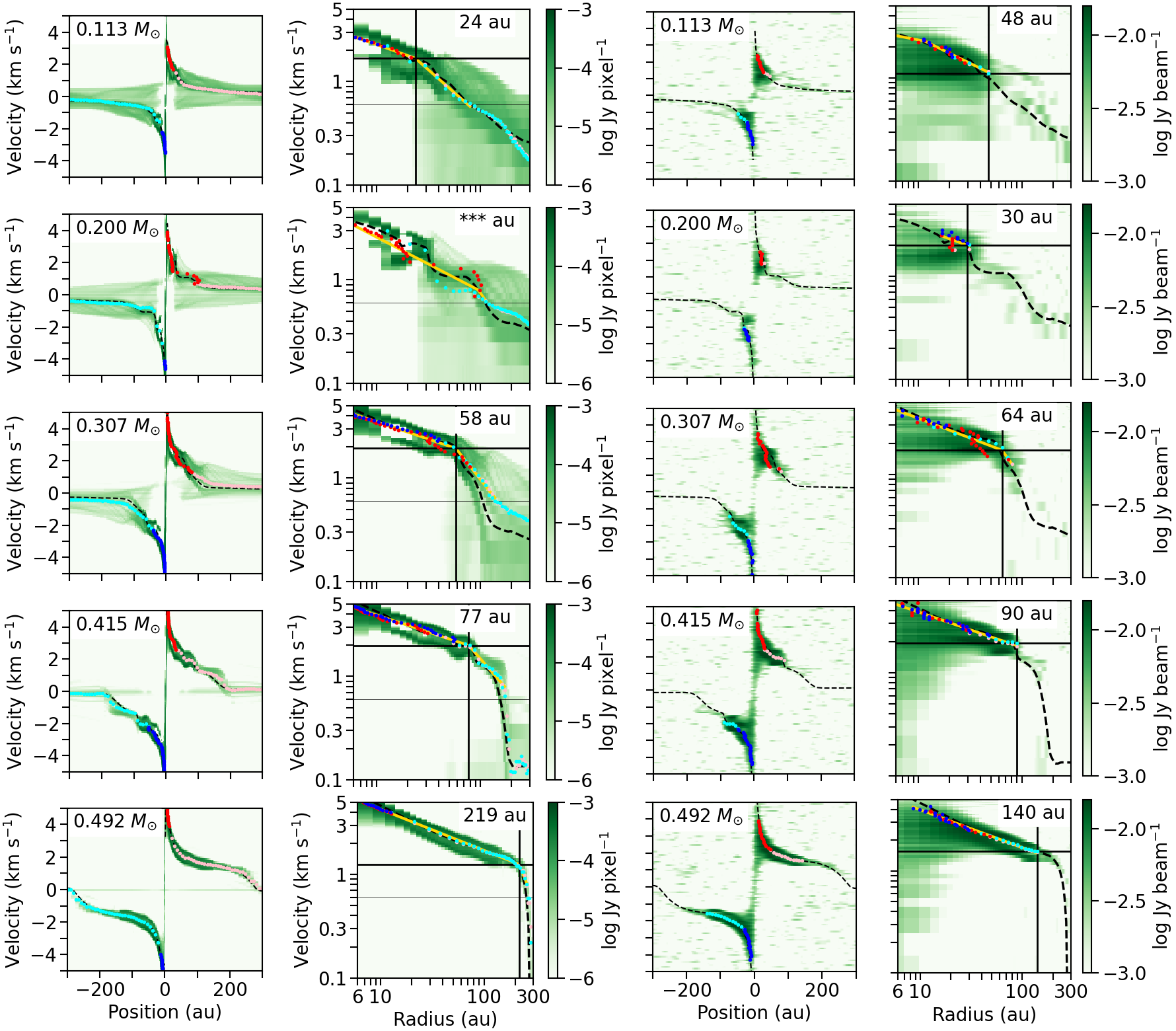}
\caption{
Linear (each left) and logarithmic (each right) position velocity (PV) diagrams cut along the major axis  before ({\it Left pair}) and after ({\it Right pair}) performing the synthetic observation. 
The logarithmic diagram shows the mean intensity of the first and fourth quadrants of the linear diagram.
The epochs are the same as in Figure \ref{fig:true}.
$\phi =0\arcdeg$ is adopted. 
The central stellar  mass at each epoch is denoted in the top left corner. 
The red and blue points denote the intensity-weighted mean position at each velocity channel, while the light blue and pink points denote the intensity-weighted mean velocity at each position in half-beam steps. 
The dashed curve in each panel shows the actual rotational velocity taken from the simulation, reduced by the inclination factor. 
The black vertical and horizontal lines denote the measured disk radius and the corresponding velocity in each panel.
In the logarithmic PV diagram (each right), the yellow line indicates the best-fit double-power-law function, while the thin horizontal line corresponds to a velocity of $0.6~\kms$.
\label{fig:pv}}
\end{figure*}

Figure \ref{fig:pv} shows linear and logarithmic PV diagrams along the major axis before and after performing the synthetic observation. 
The mean intensity of the first and fourth quadrants in the linear diagrams is plotted in the logarithmic diagrams. 
We can confirm a spin-up rotation from the PV diagrams, in which high velocity emission appears near  the center. 
At the epoch of $M_*=0.492~\Ms$ before performing the synthetic observation (bottom left two panels of Figure~\ref{fig:pv}), the velocity rapidly decreases around  $\pm 200$--300 au, which corresponds to the transition layer between the counterclockwise and clockwise rotation seen in the middle column of Figure \ref{fig:true}. 
The PV diagrams after performing the synthetic observation (right two panels of Figure~\ref{fig:pv}) indicate that strong emission is compact in the early phases, while it is more extended in the later (or evolved) phases. 
This trend can be explained by the disk growth as seen in Figure \ref{fig:nT}.

In Figure \ref{fig:pv}, the mean positions are plotted as blue and red points, while the mean velocities are light blue and pink points. 
These representative points are derived down to $\sim0.2~\kms$ before performing the synthetic observation.
On the other hand,  the points are derived only down to $~\sim 1.5~\kms$ after performing the synthetic observation because of the $5\sigma$ cutoff, i.e., sensitivity limit. 
Overall, these points trace the actual rotational velocity, which is reduced by the inclination factor ($\sin i$) and is shown by the black dashed curve both before and after performing the synthetic observation. 

Several previous studies adopted outer edges of emission in PV diagrams \citep[e.g.,][]{alve17,mare20} which correspond to the hypotenuse of the largest triangle (contour) in Figure \ref{fig:pvm}.
The panels with the synthetic observation (two right-side panels of Figure \ref{fig:pv}) indicate that the outer edge of the emission, such as 5$\sigma$ contour, is located outside the actual rotational-velocity curve. 
This analysis thus demonstrates that the outer edge is less accurate than the peak points for tracing the rotation (or Keplerian) curve.

The power-law index of the rotational velocity as a function of the radius can be more clearly confirmed  in the logarithmic diagram (Figure~\ref{fig:pv} second and fourth columns) than in the linear diagram (Figure~\ref{fig:pv} first and third columns). 
The curve taken directly from the simulation shows a power-law index of approximately $-0.5$ in the inner region at all epochs. 
On the other hand, the curve in the outer region is not simple. 
At the epochs of  $M_*=0.113$ and $0.200~\Ms$, the power-law index is close to $-0.5$ in the range of  $r \lesssim  100$ au, which is  larger than the disk radii directly estimated from the simulation, $R_{\rm rot}$, $R_{\rm inf}$, and $R_{\rm den}$. 
In addition, the true curve shows a wavy profile at the epoch of $M_*=0.200~\Ms$, at which a strong spiral appears. 
At the epochs of  $M_*=0.415$ and $0.492~\Ms$, the inner region with a power-law index $\sim -0.5$ can be more easily distinguished from the outer region because of the sudden drop of the rotational velocity. 

To determine the range or radius applicable to the Keplerian rotation, we fit the mean points with the double power-law function:
\begin{eqnarray}
v = v_b \left(\frac{r}{r_b}\right) ^{-p}\ \ {\rm if}\ \ r<r_b\ \ {\rm else}\ \ v_b\left(\frac{r}{r_b}\right)^{-(p + dp)},
\label{eq:broken}
\end{eqnarray}
where $r$, $v$, $(r_b,v_b)$, $p$, and $dp$ are the radius, velocity (variable),  break point radius and velocity,  power-law index in the inner region, and difference of index  between the inner and outer regions, respectively. 
Since spin-up rotation is expected on the disk and envelope scales, this fitting only includes velocities higher than that of the largest radius. 
Because the rotational velocity at a radius larger than the Keplerian disk is supposed to have a steeper index (larger $-d\ln v/d\ln r$), the range of $p$ and $dp$ is limited to positive values. 

The fitting is done on the logarithmic plane, and the weighting of each point is uniform. 
If the best-fit outer index $p + dp$ is close to the Keplerian law ($\leq 0.5$), it is considered that the whole fitted range is in Keplerian rotation.
Hence, the radius having the lowest velocity is adopted as the disk radius  $R_{\rm pv}$. 
If the best-fit inner index $p$ is close to the Keplerian law but the outer index is not, the break radius $r_b$ is adopted as $R_{\rm pv}$. 
If the index $p$ is not close to the Keplerian law, $R_{\rm pv}$ cannot be defined from this analysis. 
It should be noted that, at the epochs  of $M_*=0.307$ and $0.415~\Ms$, the representative points before performing the synthetic observation, as well as the true curve, show three power-law indices: a Keplerian-like index, a steeper index, and another shallower index from the inside to the outside. 
The outermost index is not related to the boundary between the disk and  the surrounding envelope. 
To ignore the outermost  part, the representative points at velocities higher than $0.6~\kms$ are used in the fitting. 
On the other hand, the outermost part has no representative point after performing the synthetic observation because of the sensitivity cutoff.

The best-fit function is plotted by the yellow line in the logarithmic diagrams of Figure~\ref{fig:pv}. 
The derived $R_{\rm pv}$ is shown by the vertical black line and noted in the top right corner in the logarithmic diagrams. 
Overall, the best-fit function traces  the mean points well and reproduces the break of the power-law between the inner Keplerian-like part and the outer envelope-like part both before and after performing the synthetic observation. 
At the epoch of $M_*=0.200~\Ms$, $R_{\rm pv}$ is not defined because the derived power-law index is steeper than the Keplerian law even for inner radii. 
One reason for this is the wavy pattern of the true curve (or velocity profile) due to the existence of the spiral arms. In fact, this radius is defined in the case with $\phi=90\arcdeg$.

The break point $(r_b, v_b)$ can provide an estimate of the protostellar mass $M_*$, as $r_bv_b^2 / G / \sin ^2 i$, where $i$ is the inclination angle.
Meanwhile, if the best-fit power-law index is not exactly $-0.5$, the inner radii provide different masses. 
Hence, we calculate the velocity $v_i$ from the radius of the innermost representative point $r_i$ using the best-fit curve (eq.~\ref{eq:broken}), and then the mass is estimated as  $r_iv_i^2 / G / \sin ^2 i$. 
The mass $M_{\rm *pv}$, defined as the geometric mean of the two masses, will be compared with the actual protostellar mass $M_*$ in \S\ref{sec:mass}.


\subsection{Rotational Velocity and the defined Radii}
\label{sec:definition}
\begin{figure}[htbp]
\epsscale{0.75}
\plotone{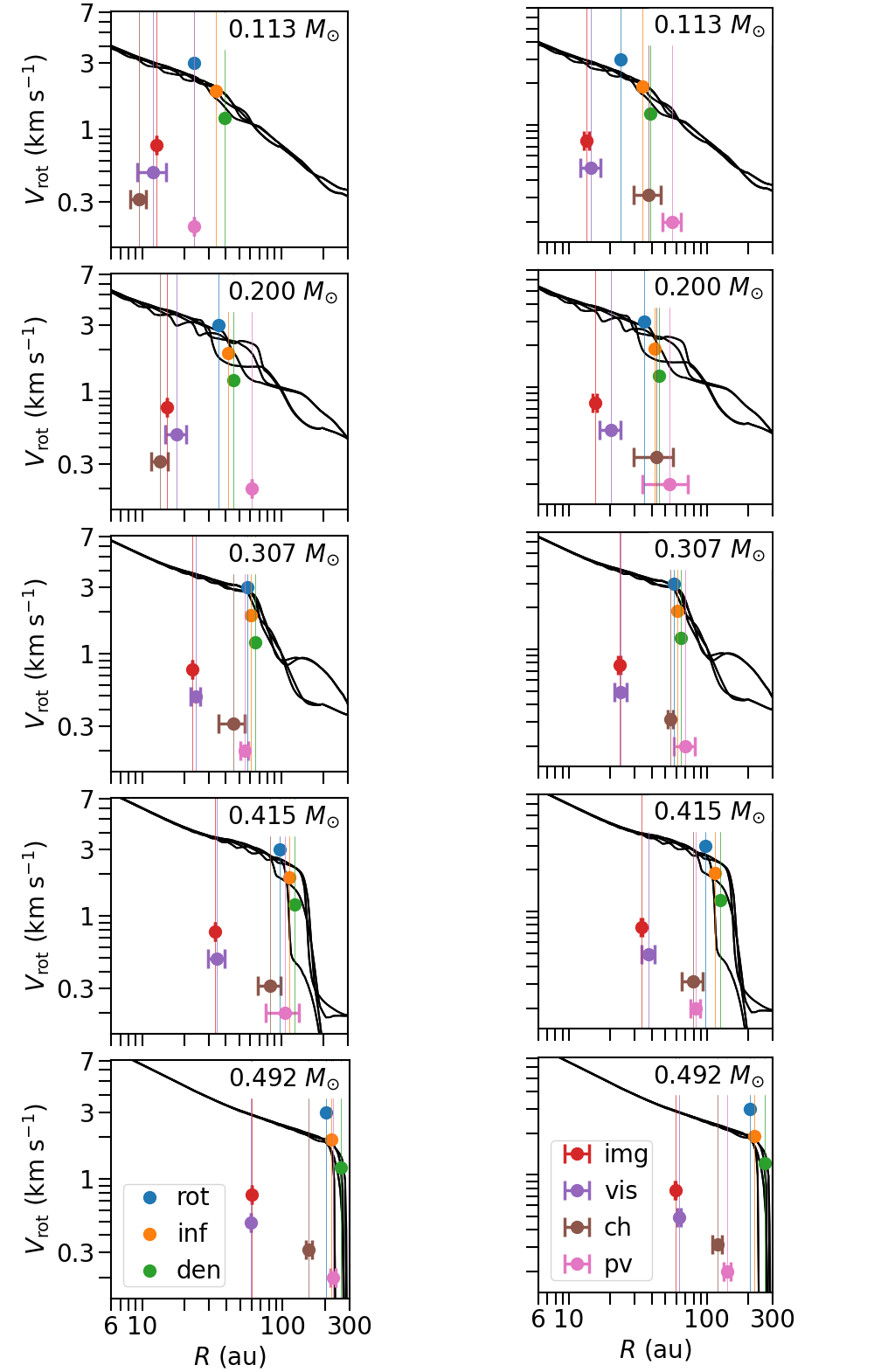}
\caption{
Seven radii $R_{\rm rot}, R_{\rm inf}, R_{\rm den}, R_{\rm img}, R_{\rm vis}, R_{\rm ch}$, and $R_{\rm pv}$ before (left panels) and after (right panels) performing the synthetic observations. 
The radii in the vertical scale is arbitrary; they are placed just to easily distinguish each other.
The velocity profiles with different azimuthal directions $0$, 45, 90 and $135\arcdeg$ are also plotted in each panel.
The error bars are estimated from the standard deviation between $\phi=0$ and $90\arcdeg$ before performing the synthetic observation (left panels) and with an iteration among models with different artificial noise  after performing the synthetic observation (right panels).
\label{fig:rot}}
\end{figure}

We defined seven radii to investigate the `disk radius'  in this study. 
It is useful  to compare the radii derived in different criteria or methods. 
The radii at different epochs are plotted in Figure \ref{fig:rot}.
The radial profiles of the rotational velocity with different azimuthal directions are also plotted in the figure.
A significant difference in the radial profiles at the same epoch means the appearance of a clear spiral. 
For example, a noticeable difference in the azimuthal variations can be confirmed in the velocity profile at the epochs of $M_* =0.02$ and $0.415~\Ms$.
At the same epoch, we can see a prominent spiral structures in the left panels of Figure~\ref{fig:nT}. 

In Figure~\ref{fig:rot}, the velocity profile can be divided into two parts.
The profile tends to show a shallow power-law index of $\sim -1/2$ in an inner part (or smaller radius), while it has a steep index in an outer part  (or larger radius).
Thus, there is a clear boundary between the inner and outer parts, in which  
 the outer power-law index is less steep in earlier epochs than in later epochs.
Note that the position of the boundary at the same epoch has a relative uncertainty of $\sim 20\%$ when the system has a clear spiral because of the azimuthal variation of the disk morphology (see Fig.~\ref{fig:nT}).

The disk is usually defined as the region where the centrifugal force is balanced with the gravitational force. 
The power-law index steeper than $-1/2$ indicates that the outer region is not supported by rotation  and thus corresponds to the infalling envelope.
The position of the bent in the velocity profile corresponds to the boundary between the rotationally supported disk and the infalling envelope, at which the power-law index of the velocity profile is changed.   
The three radii defined directly from the simulation, $R_{\rm rot}, R_{\rm inf}$, and $R_{\rm den}$, trace the disk-envelope boundary well in all epochs in Figure \ref{fig:rot}. 
Particularly, $R_{\rm rot}$ was  defined by the balance between the centrifugal and gravitational forces in \S\ref{sec:true}.
Thus, we should use $R_{\rm rot}$ to more correctly identify the disk. 
For these reasons, we regard $R_{\rm rot}$ as the actual disk radius and basically compare it with the radii defined in observational quantities.

Figure \ref{fig:rot} indicates that there is no significant difference among the radii derived directly from the simulation ($R_{\rm rot}$, $R_{\rm inf}$ and $R_{\rm den}$). 
The figure also shows that the radii defined from the continuum emission, $R_{\rm img}$ and $R_{\rm vis}$, are 2-3 times smaller than $R_{\rm rot}$ at each epoch.
On the other hand,  the radii defined from the C$^{\rm 18}$O line emission, $R_{\rm ch}$ and $R_{\rm pv}$, better trace the actual disk radius $R_{\rm rot}$, in particular, after performing the synthetic observations. 
Properties  of these observable radii will be discussed in more detail in the next section (\S\ref{sec:dis}).

\clearpage

\section{Discussion} \label{sec:dis}
Our simulation covers the evolution of a protostellar system in the range of $M_*=0$ to $0.5~\Ms$, which allows the inspection of the evolution of observable quantities as well as physical quantities.
In this section, we describe the characteristics of the disk radii defined in \S\ref{sec:ana} and then discuss the protostellar and disk masses.
We show these quantities as a function of the central stellar  mass  $M_*$ roughly every $0.02~\Ms$.
Note that the increment of mass in our simulation is much smaller than $0.02~\Ms$.

\subsection{Comparison of Radii through Protostellar Evolution} \label{sec:evo}
We defined seven different  radii,  $R_{\rm rot}$, $R_{\rm inf}$,  $R_{\rm den}$,  $R_{\rm img}$, $R_{\rm vis}$, $R_{\rm ch}$, and $R_{\rm pv}$,  in \S\ref{sec:ana}. 
Three of them, $R_{\rm rot}$, $R_{\rm inf}$, and $R_{\rm den}$, are directly derived from the simulation. 
In more detail, the centrifugal force is balanced with the gravitational force at $R_{\rm rot}$, the rotation velocity much dominates the radial velocity within $R_{\rm inf}$, and the gradient ($d\ln \rho/ d\ln r$) is the largest at $R_{\rm den}$.

\begin{figure*}[htbp]
\epsscale{1.15}
\plotone{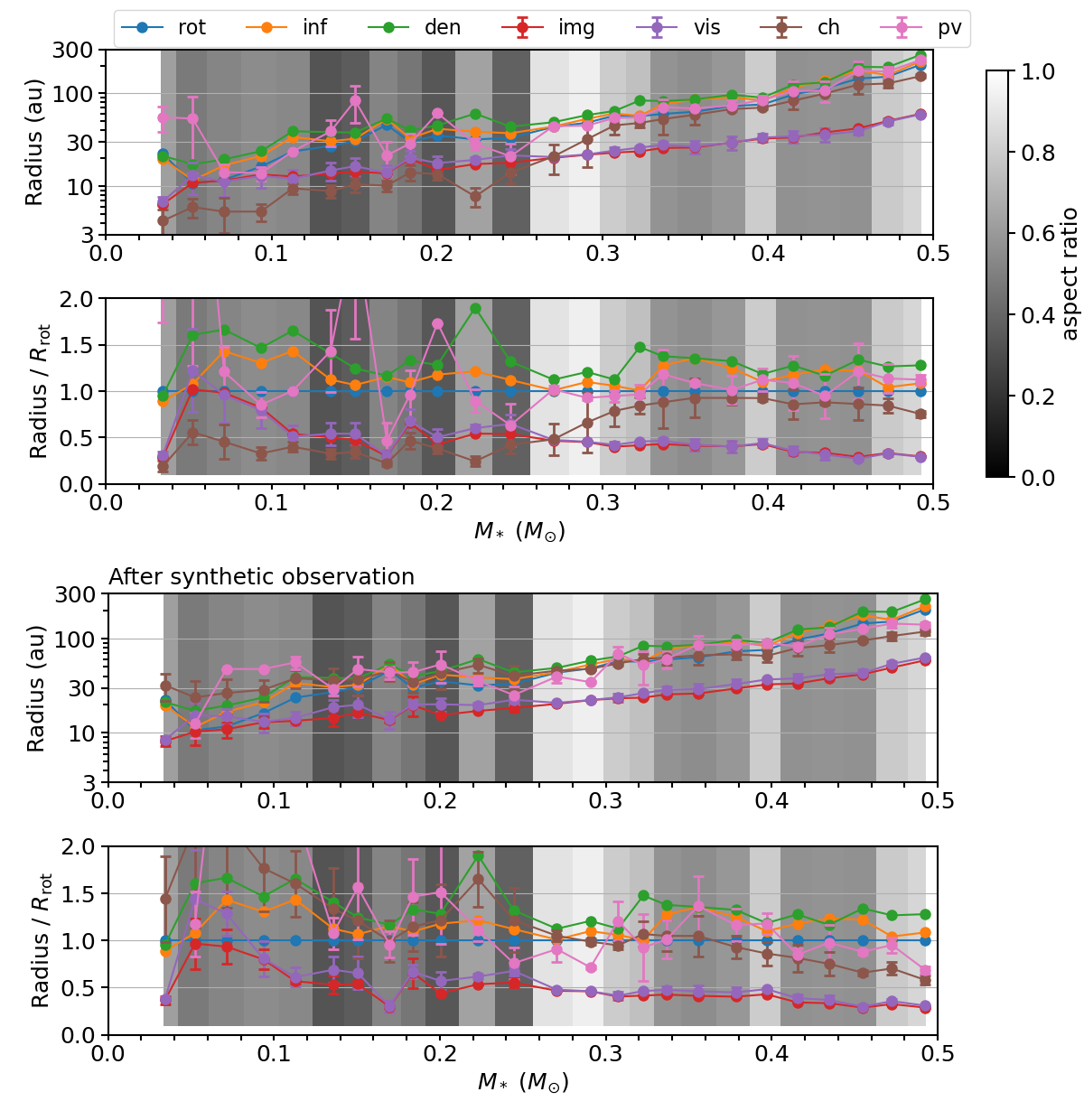}
\caption{
Evolution of the seven different radii averaged between $\phi =0$ and $90\arcdeg$. 
The upper and lower pairs are  the results before and after performing the synthetic observation, respectively. 
Each pair shows the radii (each upper panel) and the ratios of the radii to $R_{\rm rot}$ that is defined from the balance between the gravitational and centrifugal forces  (each lower panel). 
The error bars are calculated from the standard deviation between $\phi=0$ and $90\arcdeg$ (before performing the synthetic observation) and with an iteration among models with different artificial noise  (after performing the synthetic observation). 
The gray background represents the aspect ratio of the disk at each epoch, which is defined from the azimuthal variation of the three radii $R_{\rm rot}$, $R_{\rm inf}$, and $R_{\rm den}$. 
A circular morphology without a clear spiral is realized with a high aspect ratio (close to unity).
\label{fig:radii}}
\end{figure*}

The other four originate in observed maps, in which $R_{\rm img}$ corresponds to the HWHM derived from Gaussian fitting to the continuum image, $R_{\rm vis}$ is calculated from the first null point in the continuum visibility profile, $R_{\rm ch}$ is the largest radius within which 2D mean positions on the C$^{18}$O channel maps are along the major axis, and $R_{\rm pv}$ is the outermost radius showing the Keplerian law derived from 1D mean positions and velocities in the C$^{18}$O PV diagram along the major-axis.
Figure \ref{fig:radii} shows these radii before and after performing the synthetic observation as a function of protostellar mass $M_*$. 
In the figure, the error bars for the radii defined by the observable quantities ($R_{\rm img}$, $R_{\rm vis}$, $R_{\rm ch}$, and $R_{\rm pv}$)  are calculated both from  the standard deviation between $\phi=0$ and $90\arcdeg$ and from an iteration among models with different artificial noise.
The latter effect (the error due to the artificial noise) is not included in the radii without  the synthetic observation (upper two panels). 

\subsubsection{Disk Aspect Ratio}
As seen in the density distribution (Figure~\ref{fig:nT}), the disk in our simulation sometimes shows a spiral structure, and the strength (or amplitude) of the spiral varies through the simulation. 
As an indicator of the spiral strength, we define the aspect ratio using the azimuthal variation of $R_{\rm rot}$, $R_{\rm inf}$, and $R_{\rm den}$ in order to investigate the relation between the strength of the spiral and the disk radius. 
As described in \S\ref{sec:definition}, we regard $R_{\rm rot}$ as the actual disk radius, because it is defined from the viewpoint of the balance  between between centrifugal and gravitational forces and well agrees with the radial profile of rotational velocities. $R_{\rm inf}$ and $R_{\rm den}$ also trace this actual radius well.
Figure \ref{fig:true} shows that these radii have azimuthal variations and the variations become large as the spiral becomes prominent.  
For the mean and standard deviation of a radius, $m$ and $s$, respectively,  the value $(m - s) / (m + s)$ provides the ratio between the narrowest and widest sizes of the disk. 
We also define the aspect ratio as the mean of the ratios among the three radii. 

The aspect ratio is shown by the gray scale in the background of Figure \ref{fig:radii}.
An aspect ratio close to unity means that the disk has a nearly circular morphology with a weak  spiral. 
There is a clear trend in the aspect ratio through the protostellar evolution in that the aspect ratio is larger in the early half stage of the evolution,  $M_* \lesssim 0.25~\Ms$, than in the later half stage,   $M_* \gtrsim 0.25~\Ms$. 
In other words, a clear spiral  (or a distorted disk) tends to appear in the early stage, while a circular disk can be seen in the later stage. 

\subsubsection{Disk Radii Estimated from Simulation}
As described above, the radii $R_{\rm inf}$ and $R_{\rm den}$ are supposed to well trace the actual disk radius $R_{\rm rot}$. 
For $M_* \lesssim 0.15~\Ms$,  the radius $R_{\rm rot}$ is the smallest among the three radii, with  $R_{\rm inf} \sim 1.4R_{\rm rot}$ and  $R_{\rm den}\sim  1.6R_{\rm rot}$. 
On the other hand, for  $M_* \gtrsim 0.15~\Ms$, $R_{\rm inf}$ and $R_{\rm den}$ are close to $R_{\rm rot}$; both $R_{\rm inf}$ and  $R_{\rm den}$ are about 30\% larger than $R_{\rm rot}$ ($\sim 1.3R_{\rm rot}$).
Note that there is an exception even for  $M_* \gtrsim 0.15~\Ms$; $R_{\rm den}$ is twice as large as $R_{\rm rot}$ and $R_{\rm inf}$ at the epoch of  $M_* \simeq0.22~\Ms$. 
The relatively large $R_{\rm inf}$ and $R_{\rm den}$ compared with $R_{\rm rot}$ for $M_* < 0.15~\Ms$ suggest a wide range of the transition layer between the rotating disk and surrounding envelope, as seen in Figure~\ref{fig:rot}. 
For $M_*<0.15~\Ms$, the infall velocity slowly diminishes and the density slowly rises as the distance from the central star decreases because the gravity of the protostar is not very strong.

The difference between the radii  $R_{\rm rot}$, $R_{\rm inf}$, and $R_{\rm den}$ for $M_* < 0.15~\Ms$ also implies that the disk  can have an intrinsic uncertainty in its size of as much as $\sim 60\%$, depending on which quantity (rotation, infall motion, or density) is used  to estimate the disk radius, regardless of the analysis method in the observations. 
Although the uncertainty in the disk size is smaller  for  $M_* \gtrsim 0.15~\Ms$ than for  $M_* \lesssim 0.15~\Ms$,  it is still $\sim 20$--30\% even for the later phase.

\subsubsection{Disk Radii Estimated from Continuum Emission}
The radii $R_{\rm img}$ and $R_{\rm vis}$ derived from the continuum emission behave very similarly, even though the different distributions of Gaussian  ($R_{\rm img}$)  and boxcar ($R_{\rm vis}$) disks were assumed to estimate them.  
The sizes of the two radii are also the same both before and after performing the synthetic observation. 
This is because the continuum emission is detected at a $> 1000\sigma$ level for the emission peak, and the beam size of 20~au is similar to or smaller than the diameters of the disk. 
This high sensitivity and high angular resolution reduces the difference between them.

The visibility-radius $R_{\rm vis}$ after performing the synthetic observation is noticeably larger than the image-radius $R_{\rm img}$ at the epochs of $M_*=0.05$, 0.13, and $0.20~\Ms$. 
This can be attributed to the development of the spiral stricture,  as discussed in \S\ref{sec:vis}.
When estimating $R_{\rm vis}$, either $\phi=0\arcdeg$ or $90\arcdeg$ can have a closer null point than the other directions of  $\phi$, and  the long axis is preferentially adopted as $R_{\rm vis}$.
Thus,  $R_{\rm vis}$ becomes larger than $R_{\rm img}$ when a clear spiral appears. 

As seen in Figure \ref{fig:radii}, the continuum radii ($R_{\rm img}$ and $R_{\rm vis}$) do not correctly trace the actual disk  radius $R_{\rm rot}$.
They are always smaller than the radii derived from the simulation.
Although the continuum radii $R_{\rm img}$ and $R_{\rm vis}$ are barely consistent with the actual radii for $M_* \lesssim 0.1~\Ms$, they are roughly 0.3 times as small as the actual radii for $M_* \gtrsim 0.45~\Ms$. 
The underestimation of these continuum radii indicates that the fitting to continuum images depends more strongly on the central peak structure than on the disk outer edge. 
As the disk grows, the disk outer edge moves outward and the  density and temperature near the outer edge decrease, which causes the emission from the outer disk region to be fainter and harder to be detected. 
Therefore, the radii $R_{\rm img}$ and $R_{\rm vis}$ are considerably smaller than the actual radius $R_{\rm rot}$ especially in later phases.
This can explain the decrease trend of the ratio between these continuum radii and $R_{\rm rot}$ during the whole evolution (Fig.~\ref{fig:radii}).
Figure \ref{fig:radii} may provide a correction from an observed continuum radius to the actual  disk radius as a function of $M_*$ when $M_*$ is estimated.

\subsubsection{Disk Radii Estimated from Line Emission}
The radius derived from the C$^{18}$O channel maps, $R_{\rm ch}$, appears to closely trace the actual disk radius for $M_* > 0.3~\Ms$. 
However, this could be a coincidence, particularly before performing the synthetic observation, as discussed in \S\ref{sec:mass}. 
For $M_* < 0.3~\Ms$, the radius $R_{\rm ch}$ is smaller/larger than the actual disk radius before/after performing the synthetic observation. 
The estimation of  $R_{\rm ch}$ depends on the lowest velocity included in the emission from the disk.
Such a low velocity can, however, also include emissions from the surrounding envelope, which shifts the velocity gradient from the major axis to the minor axis. 
The underestimation of $R_{\rm ch}$ in the early phase ($M_* < 0.3~\Ms$) before performing the synthetic observation can  be attributed  to the significant envelope emission comparable to the disk emission. 
On the other hand, the synthetic observation resolves out the envelope, making the emission region more compact, as seen in the integrated intensity maps (Fig.~\ref{fig:chan}).
In this case, the emission peak is aligned more closely to  the major axis. 
This can explain the overestimation of $R_{\rm ch}$  in the early phase ($M_* < 0.3~\Ms$) after performing the synthetic observation.  
In addition, the radius $R_{\rm ch}$ tends to be slightly smaller than the actual disk radius, particularly after performing the synthetic observation for $M_* \gtrsim 0.4~\Ms$, which can be explained by the contrast of emissions between the inner and outer parts. 
The emission contrast between inner and outer radii is higher  in the later phase than in the early phase, because the disk outer edge is located far from the center and thus the emission is weak.
Such emission also tends to be resolved out.
This high contrast shifts the mean position more inward than our correction \citep[Appendix A of ][]{aso15} assuming a uniform emission in a disk.

The radii derived from the C$^{18}$O PV diagrams, $R_{\rm pv}$, agree well with the actual disk radius when $M_* \gtrsim 0.2~\Ms$ both before and after performing the synthetic observation. 
In the early phase, $R_{\rm pv}$ is larger than the actual disk radius  because the true rotational velocity does not drop rapidly outside the disk, i.e., in the envelope at the epoch of $M_*\lesssim 0.1$--$0.2~\Ms$, as seen in \S\ref{sec:pv}. 
In addition, the envelope is also more massive in the early phase than in the later phase. 
As a result, in the early phase,  it is difficult to distinguish the mean positions of the envelope rotation from that of the Keplerian rotation, which results in overestimation of the disk radius.

In the synthetic observation, $R_{\rm pv}/R_{\rm rot}$ tends to decrease for $M_*\gtrsim 0.35~\Ms$ and is noticeably smaller than the actual disk radius at the epoch of $M_* \sim 0.5~\Ms$. 
This can be explained by the sensitivity limit,  because the emission from the disk outer edge is weak when the disk is evolved and has a large size.
In other words, if all of the representative points can be explained by the Keplerian law, these points provide only a lower limit for the disk radius.

\subsection{Estimation of Protostellar Mass} \label{sec:mass}
In addition to the disk radius, the Keplerian disk includes important information on the central stellar mass $M_*$, which is one of the most reliable indicators of stellar evolution. 
The velocity distribution or line observation is required to estimate $M_*$. 
Among the four radii $R_{\rm img}$, $R_{\rm vis}$, $R_{\rm ch}$, and $R_{\rm pv}$ that can be estimated from observational data, $R_{\rm ch}$ and $R_{\rm pv}$ are estimated from  line observations (or channel map and PV diagram). 
Thus, they can provide an estimation of $M_*$ ($M_{\rm *ch}$ and $M_{\rm *pv}$), as defined in \S\ref{sec:chan} and \S\ref{sec:pv}.

\begin{figure}[htbp]
\epsscale{0.8}
\plotone{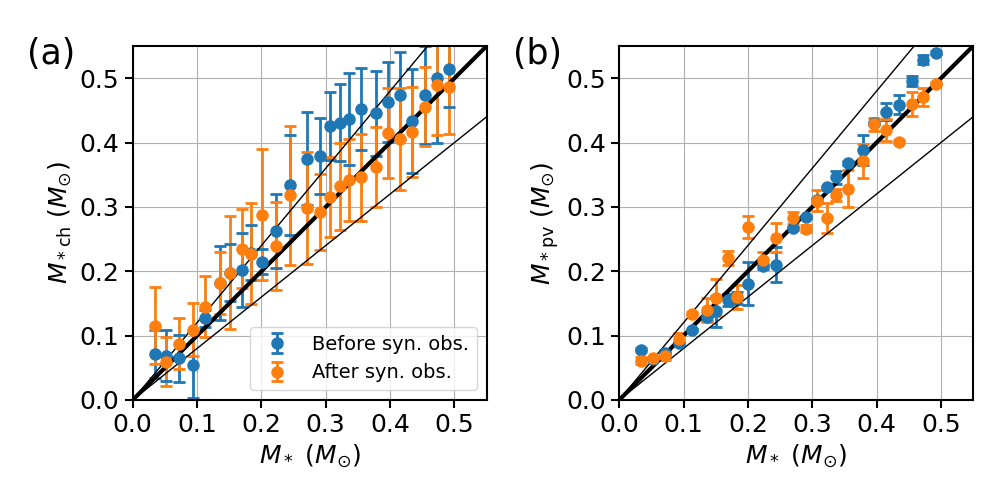}
\caption{
Central stellar mass estimated from the C$^{18}$O channel maps $M_{\rm *ch}$ (left)  and PV diagrams $M_{\rm *pv}$ (right)  as a function of the central stellar  mass $M_*$. 
The former  is estimated to be $M_{\rm *ch}=R_{\rm ch}v_{\rm ch}^2/G/\sin ^2 i$, where $v_{\rm ch}$ is the velocity of the outermost mean position along the major axis.
The latter mass $M_{\rm *pv}$ is the geometric mean of $r_b v_b ^2 / G / \sin ^2 i$ and $r_i v_i ^2 / G / \sin ^2 i$, where $(r_b, v_b)$ is the break point of the logarithmic PV diagram and $(r_i, v_i)$ is the innermost point of the best-fit double power-law function. 
The blue and orange points are the masses before and after performing the synthetic observation, respectively. 
The uncertainties come from the standard deviation between $\phi=0$ and $90\arcdeg$  and that between the iterations with artificial noise. The three solid lines in each panel correspond to $M_{\rm *ch}$, $M_{\rm *pv} = $ 0.8, 1.0, and $1.2M_*$, respectively.
\label{fig:mchpv}}
\end{figure}

Figure \ref{fig:mchpv} shows $M_{\rm *ch}$ and $M_{\rm *pv}$ as a function of $M_*$. 
Before performing the synthetic observation, $M_{\rm *ch}$ is larger than $M_*$ in the range $0.25~\Ms \lesssim M_* \lesssim 0.4~\Ms$, although their errors are as large as $\gtrsim 0.05~\Ms$. 
After performing the synthetic observation, $M_{\rm *ch}$ is roughly in agreement  with $M_*$  within $\pm 20\%$ accuracy, although their errors are also as large as $\gtrsim 0.1~\Ms$. 
Multiple factors could shift $M_{\rm *ch}$ up or down in the panel. 
For example, the disk self-gravity increases  $M_{\rm *ch}$ because it is estimated from the rotational velocity, which is determined by the sum of the disk self-gravity and point gravity of the central star. 
On the other hand, the intensity contrast between the inner and outer regions in the disk can shift the 2D mean positions inward and consequently lead to underestimation of  $M_{\rm *ch}$. 
Our correction does not work very well  with high contrast, because the correction assumes an uniform intensity distribution in the disk, as mentioned in \S\ref{sec:chan}.

The mass estimated from the PV diagram $M_{\rm *pv}$ is in good agreement with $M_*$ for all epochs both before and after performing the synthetic observation. 
Furthermore, their errors are significantly smaller than those for $M_{\rm *ch}$.
Only emissions from the major axis are extracted from the PV diagrams.
Thus, the 1D mean points in the PV diagrams are less affected by the emissions not along the major axis, such as the emissions from the envelope, unlike the case of the 2D mean positions adopted in channel maps. 
Therefore, $M_{\rm *pv}$ traces $M_*$ better than $M_{\rm *ch}$. 
This trend is more clear before performing the synthetic observation, where a large amount of  emission from the envelope makes the disk-envelope boundary clearer. 
It should be noted that $M_{\rm *pv}$ is slightly larger than the actual mass when $M_* < 0.4~\Ms$. 
This is caused by fact that in an early phase the disk mass contributes to the rotational velocity that determines the central stellar mass.
Meanwhile, after performing the synthetic observation, $M_{\rm *pv}$ is estimated  from a radius inside the outermost radius of the disk  because of the sensitivity cutoff, which tends to derive a value of $M_{\rm *pv}$ similar to the actual stellar mass $M_*$.

\subsection{Effects of Inclination Angle} \label{sec:incl}
Assuming an inclination angle of $i=45\arcdeg$, the radiative transfer process is calculated to derive the observational quantities in \S\ref{sec:ana}. 
In this subsection, we investigate the effect of the inclination angles $i$ on the estimation of the disk radius and central stellar mass. 

\begin{figure*}[htbp]
\epsscale{1.175}
\plotone{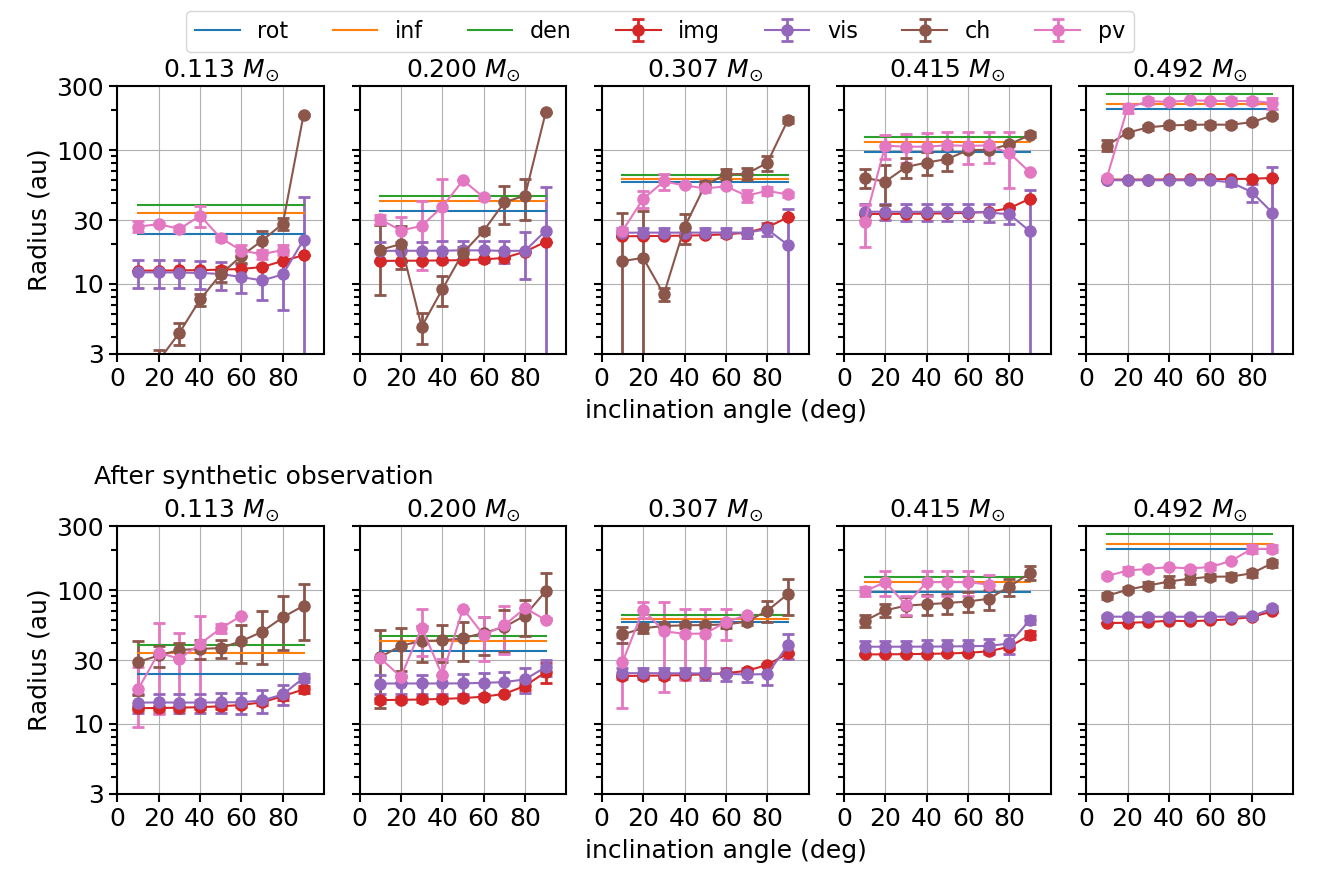}
\caption{
Effects of inclination angles on the estimation of disk radii before (top) and after (bottom) performing the synthetic observation at different epochs. 
The three lines without symbols (`rot', `inf', and `den') are taken from the simulation and independent of the inclination angle. 
The central stellar  mass is denoted at the top of each panel. 
\label{fig:rincl}}
\end{figure*}

Figure \ref{fig:rincl} shows the estimated disk radii with different inclination angles $i$ (every $10\arcdeg$ steps) at five different epochs.
The actual disk radii directly derived from the simulation, $R_{\rm rot}$, $R_{\rm inf}$, and $R_{\rm den}$, are independent of $i$. 
Regardless of whether or not the synthetic observation is performed, the continuum radii  $R_{\rm img}$ and $R_{\rm vis}$ have the same trend, that is,  they are smaller than the actual disk radii $R_{\rm rot}$ independent of $i$. 
Note that the edge-on case $(i=90\arcdeg)$ is an exception, in which the continuum radii ($R_{\rm img}$ and $R_{\rm vis}$) are somewhat larger in the $i=90\arcdeg$ case than  in the $i\ne90\arcdeg$ cases. 
The edge-on case exception is due to the large column density along the LOS at a given radius. 

Before performing the synthetic observation, the radius derived from the channel map $R_{\rm ch}$  strongly depends on the inclination angle $i$, especially at the epochs of $M_*=0.113$, 0.200, and $0.307~\Ms$. 
When the inclination angle is close to $i=0\arcdeg$ (face-on case), the rotational velocities in the disk are squeezed to low LOS velocities and do not significantly affect the determination of $R_{\rm ch}$. 
In addition,  the density and velocity of the envelope are spherically distributed.
Thus, for the face-on case, the envelope velocity dominates the disk rotational velocity and tends to determine $R_{\rm ch}$.
In contrast, when the system is close to $i=90\arcdeg$ (edge-on case), most of the disk component is distributed along the disk major axis. 
On the other hand, the infall motion of the envelope cannot efficiently shift the 2D mean positions off from the major axis and thus the contribution from the envelope emission is limited. 
Therefore,  as seen in Figure \ref{fig:rincl}, $R_{\rm ch}$ is large for large $i$ and small for small  $i$.
The inclination effect is stronger in the early phase, because the effect of the infalling envelope is significant. 
The same trend can be confirmed  after performing  the synthetic observation.
However,  the dependence on $i$ after performing  the synthetic observation is weaker than before performing the synthetic observation, particularly at the epochs of $M_*=0.113$, 0.200, and $0.307~\Ms$, because the emission from the envelope is partly resolved out and the effect of the envelope is not significant.

When the radius defined from PV diagrams $R_{\rm pv}$ is derived, the radius traces the actual radii $R_{\rm rot}$ better than the other radii ($R_{\rm img}$, $R_{\rm vis}$, $R_{\rm ch}$) at all the five epochs with any inclination angle, both before  and after performing the synthetic observation. This is consistent with the case of $i=45\arcdeg$ (see \S\ref{sec:ana}). 
Meanwhile, at the epochs of 0.113, 0.307, and $0.415~\Ms$, the radius $R_{\rm pv}$ is not derived when the system is close to edge-on ($i\simeq 90\arcdeg$) after the synthetic observation is performed. 
This is mainly because the line emission is detected only in a narrow range of radius in these cases.

\begin{figure*}[htbp]
\epsscale{1}
\plotone{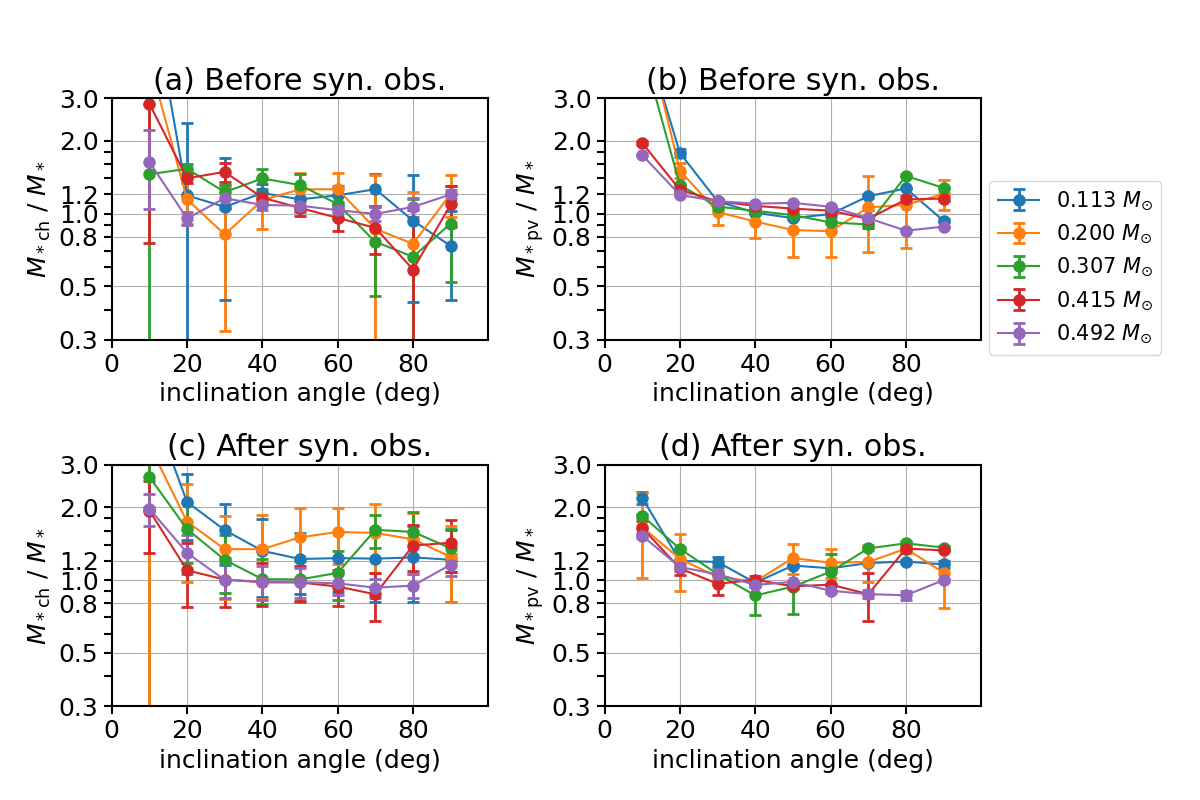}
\caption{
Effects of the inclination angle on the estimation of the central stellar mass. 
$M_{\rm *ch}$ is the mass estimated from the 2D mean position in the channel map. 
$M_{\rm *pv}$ is the mass estimated from 1D mean position/velocity in the PV diagram.
\label{fig:mincl}}
\end{figure*}

Figure \ref{fig:mincl} shows the estimation of the central stellar mass for different inclination angles $i$. 
Compared with the actual disk  mass $M_*$,  both masses estimated from the channel map ($M_{\rm *ch}$ ) and PV diagram ($M_{\rm *pv}$) are overestimated by a factor of $\sim 2$ when the inclination angle is $i=10\arcdeg$. 
This is because the infall motion in the envelope dominates the LOS velocity and the emission originating in the disk rotation is squeezed to low LOS velocities with small $i$,  as discussed above. 
With larger $i$, the ratio $M_{\rm *ch}/M_*$ ranges from $\sim 0.8$ to $\sim 2.0$.  
For the mass estimated from the PV diagram $M_{\rm *pv}$, the ratio $M_{\rm *pv}/ M_*$ ranges from $\sim 0.8$ to $\sim 1.4$, while it is suppressed within  $\sim 1.2$ with  $i\lesssim 70\arcdeg$.  
Thus,  $M_{\rm *pv}$ can trace $M_*$ better than $M_{\rm *ch}$, as expected from the analysis of the $i=45\arcdeg$ case (see \S\ref{sec:ana}).

\subsection{Gravitational Instability of Disk} \label{sec:ins}
Our simulation shows spiral arms in the disk, which suggests that the disk is massive enough to be gravitationally unstable. 
Since the disk radius can be defined directly from the simulation, the disk mass can also be calculated from the simulation. 
The three radii $R_{\rm rot}$, $R_{\rm inf}$, and $R_{\rm den}$ are defined from the simulation, in which $R_{\rm rot}$ is the smallest among them through the protostellar evolution (Figure \ref{fig:radii}) and thus provides the smallest disk mass. 
Hence, the radial range of the disk is limited to $\sqrt{x^2 + y^2}<R_{\rm rot}(\theta)$, where $R_{\rm rot}(\theta)$ is the radius before being azimuthally averaged (see \S \ref{sec:true}) and varies dependent on  the angle $\theta$. 

\begin{figure}[htbp]
\epsscale{0.5}
\plotone{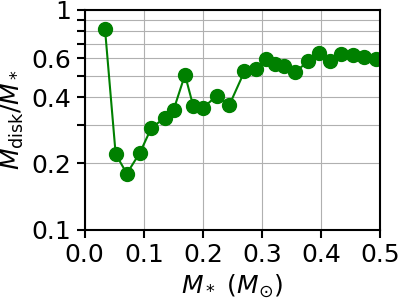}
\caption{
Ratio of the disk mass $M_{\rm disk}$ to the central stellar mass $M_*$. 
The disk mass $M_{\rm disk}$ is calculated by integrating all the mass in the region within $\sqrt{x^2 + y^2} < R_{\rm rot}(\theta)$ and $|z|<0.25\sqrt{x^2 + y^2}$, where the disk radius $R_{\rm rot}(\theta )$ varies as a function of the azimuthal angle $\theta $.
\label{fig:mdisk}
}
\end{figure}

Figure \ref{fig:nT} indicates that the vertical range of the disk is roughly within $|z|< 0.25\sqrt{x^2 + y^2}$. 
Then, we define the disk mass $M_{\rm disk}$ as the mass within these radial and vertical ranges. 
Figure \ref{fig:mdisk} shows the ratio of the estimated mass to  the central stellar mass $M_*$ as a function of $M_*$. 
Except for the earliest epoch, the ratio starts from $\sim 0.2$ and  remains $\sim 0.6$ after the central stellar mass reaches  $M_*\sim 0.3~\Ms$.
The disk instability is often investigated with the Toomre $Q$ parameter.
When a Keplerian disk and  hydrostatic equilibrium in the vertical direction are assumed,  $Q$ can be written as 
\begin{eqnarray}
Q=\frac{c_s\kappa}{\pi G\Sigma} = \frac{H\Omega ^2}{\pi G\Sigma}=\frac{H}{r}\frac{M_*}{\pi r^2\Sigma}\sim \frac{H}{r}\frac{M_*}{M_{\rm disk}},
\end{eqnarray}
where $c_s$, $\kappa$, $\Sigma$, $H$, and $\Omega$ are the sound speed, epicyclic frequency, surface density, scale height, and angular frequency of the disk, respectively. 
Our simulation shows $H/r\sim 0.25$ and $M_{\rm disk}/M_*\sim $0.2--0.6, which corresponds to $Q\sim$ 0.4--1.3. 
This suggests that the disk is in a gravitationally unstable state, which is consistent with the development of spiral arms seen in the simulation. 
\citet{tomi17} reported the spatial distribution of the $Q$ parameter in more detail.
The other two radii, $R_{\rm inf}$ and $R_{\rm den}$, which are larger than $R_{\rm rot}$, provide larger $M_{\rm disk}$ and thus smaller $Q$. 
Our simulation also includes the epochs at which the disk has a circular morphology.
This suggests that a gravitationally unstable disk can be circular temporarily and the duration of the stable state  is comparable to that of the  unstable state, which can be confirmed in Figures~\ref{fig:nT} and \ref{fig:radii}.
\citet{lo.ri05} firstly pointed out that the amplitude of the spiral arms caused by gravitational instability changes with time when the disk mass is high compared to the stellar mass. 
It should be noted that the disk size limits the growth of gravitational instability \citep{machida10,bo.cl19}. 

\section{Conclusions} \label{sec:con}
In this study, through radiative transfer and synthetic observations, we tested four observable radii by comparing with three disk radii defined directly from the simulation.
The three simulation radii were determined  according to different criteria, namely,  the rotational support ($R_{\rm rot}$), infall velocity ($R_{\rm inf}$), and  density contrast ($R_{\rm den}$). 
The four observable radii are defined from the continuum image ($R_{\rm img}$), continuum visibility ($R_{\rm vis}$), channel map ($R_{\rm ch}$), and PV diagram ($R_{\rm pv}$). 
The radiative transfer and synthetic observations mimic the observation of the 1.3 mm continuum and  C$^{18}$O $J=2-1$ line emission with a 20 au spatial resolution, a $0.1~\kms$ velocity resolution, and sensitivities of $\sigma=0.02$ and $1~\mJB$.
We obtained the following results.

\begin{enumerate}
\item The disk radius is not strictly determined even in simulations, because different schemes for the disk identification provide different disk radii. 
In the simulation,  we defined the disk using three different physical quantities, the rotation velocity $R_{\rm rot}$, infall motion $R_{\rm inf}$, and density contrast $R_{\rm den}$.
Of these,  $R_{\rm inf}$ and $R_{\rm den}$ can be $\sim 30\%$ larger than $R_{\rm rot}$ through the protostellar evolution, causing an intrinsic and non-observational uncertainty. 
    \item The methods for deriving the disk radius with the channel map ($R_{\rm ch}$) and PV diagram ($R_{\rm pv}$) can reasonably estimate the disk radius when the central stellar mass is larger than $M_* \gtrsim 0.2~\Ms$. 
Both $R_{\rm ch}$ and $R_{\rm pv}$ trace the actual disk radius better in most phases, whereas they tend to be $\sim 30\%$ smaller than the actual disk radius when $M_* \sim 0.5~\Ms$ because the sensitivity limits the disk outer radius in the later phase.
\item The observable radii $R_{\rm img}$ and $R_{\rm vis}$ underestimate the disk radius by a factor of 2--3, and the underestimation is greater in the later phase than in the early phase.  
    \item The analyses using the channel map and PV diagram also allow the central stellar mass to be estimated. 
The estimated mass is in  good agreement with the actual stellar mass through the protostellar evolution, in which the uncertainty is within $\pm 20\%$.
    \item The inclination angle $i$ does not significantly affect the estimation of the continuum radii ($R_{\rm img}$ and $R_{\rm vis}$), whereas the line radii ($R_{\rm ch}$ and $R_{\rm pv}$) more or less depends on the inclination angle.
Nevertheless, the analysis using the PV diagram still provides a reasonable estimate of the disk radius and central stellar mass  among the methods examined in this study.
    \item Spiral morphology is often seen in the disk, because the mass ratio of the disk to protostar is as large as $M_{\rm disk}/M_*= 0.2$--$0.6$ through the protostellar evolution. 
The disk also  shows circular morphology temporarily.
The periods with and without a spiral are roughly the same.
\end{enumerate}

\acknowledgments
This research used the computational resources of the HPCI system provided by the Cyber Science Center at Tohoku University, the Cybermedia Center at Osaka University, and the Earth Simulator at JAMSTEC through the HPCI System Research Project (Project ID: hp180001, hp190035, hp200004).
The  simulations reported in this paper were also performed by 2019 and 2020 Koubo Kadai on the Earth Simulator (NEC SX-ACE) at JAMSTEC. 
The present study was supported  by JSPS KAKENHI Grants (JP17H02869, JP17H06360, JP17K05387, JP17KK0096: MNM).

\vspace{5mm}
\facilities{}
\software{CASA \citep{mcmu07}, RADMC3D \citep{dull12}}

\appendix
\section{More realistic treatment of noise and continuum subtraction} \label{sec:app}
We adopted different treatment of noise inclusion and continuum subtraction from that of real interferometric observations, in order to realize the investigation of the noise effect with 100 times iterations using different noise data in a reasonable computational time. 
As described in \S\ref{sec:ana}, we subtracted the continuum emission from the image domain before performing the synthetic observation. Then, we added artificial Gaussian noise in the image domain after performing the synthetic observation (inverse-Fourier transform and CLEAN).
On the other hand, in real  observations, noise occurs in the $uv$ domain 
because visibilities are taken in interferometric observations.
The different treatment of noise inclusion and continuum subtraction may affect the results. 
In this section, we investigate whether or not these two differences affect our conclusions.  

We adopted a way of noise inclusion and continuum subtraction similar to that in real observations and executed the same analyses done in \S\ref{sec:ana} only for five representative epochs (0.113, 0.200, 0.307, 0.415, and $0.492~\Ms$). 
The radiative transfer process was calculated in all channels from $-15~\kms$ to $15~\kms$ using RADMC3D.
The output model-image cubes were passed through the CASA task {\it simobserve} to generate visibilities, i.e., the synthetic observation process. 
Then, artificial noise was added to the visibilities using the CASA task {\it sm.setnoise} with the `simplenoise' option, in which the noise amplitude was adjusted to produce $1~\mJB$ and $20~\muJB$ for the line and continuum images, respectively.  
Then, the continuum emission was subtracted through the CASA task {\it uvcontsub}, in which the continuum level was identified using the channels from $-15$ to $-12~\kms$ and 12 to $15~\kms$. 
Note that the velocity range $12~\kms <|V|<15~\kms$ was selected because the model cubes do not show emission excess from the continuum emission in this velocity range. 
This continuum subtraction provides continuum visibilities made in the range of $12~\kms <|V|<15~\kms$, as well as continuum-subtracted line visibilities. 
These visibility sets were inverse-Fourier Transformed and CLEANed to generate the final line image cube and continuum image. 
In addition, the visiblities were averaged in the time direction for the analysis of $R_{\rm vis}$ using the CASA task {\it split}. 

\begin{figure}[htbp]
\epsscale{0.8}
\plotone{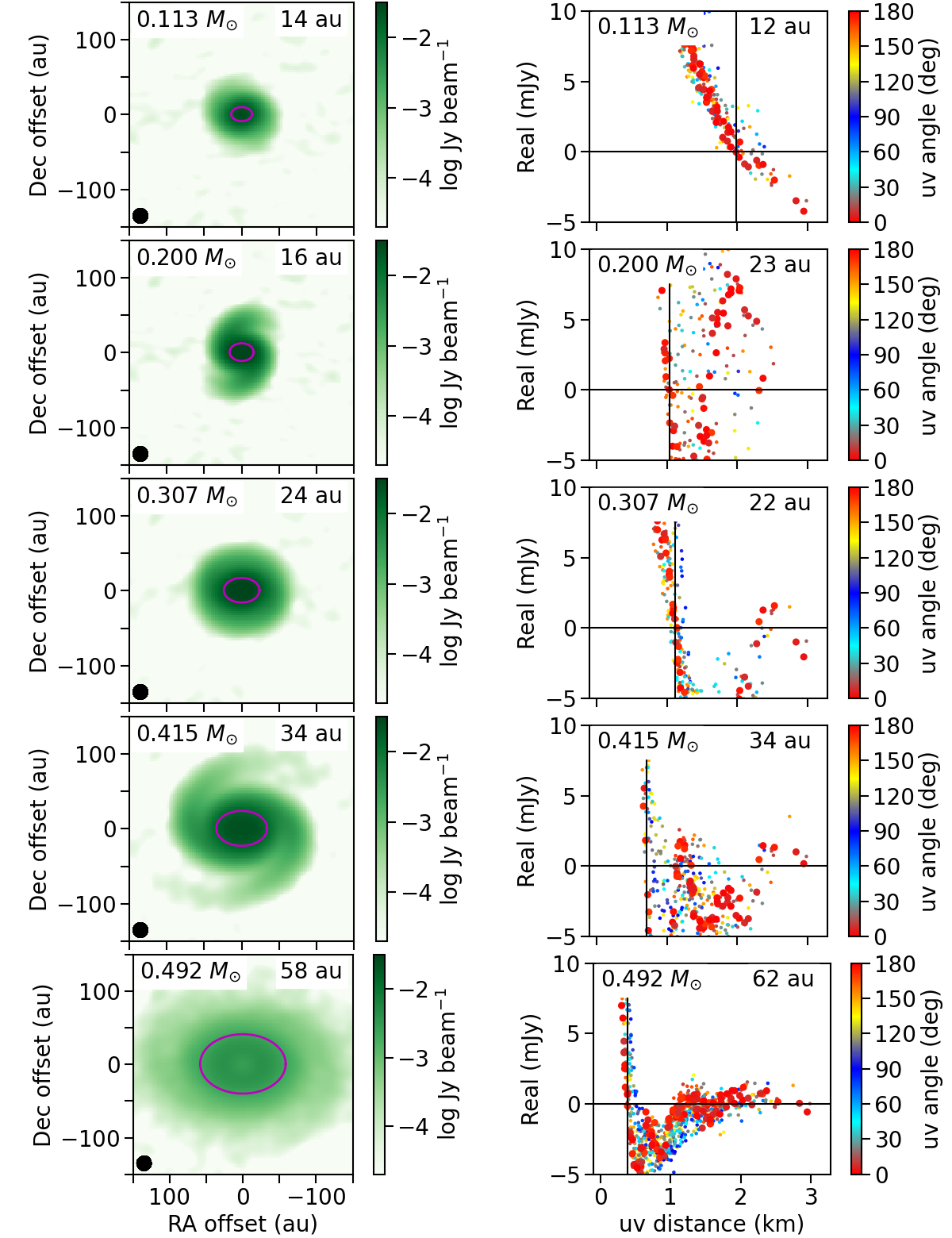}
\caption{
Same as the right panels of Figures~\ref{fig:img} and \ref{fig:vis} but with a more  realistic treatment of noise and continuum subtraction, in which the continuum visibilities are subtracted after artificial noise is added to the visibilities and then the data set is CLEANed.
\label{fig:realc}
}
\end{figure}

\begin{figure}[htbp]
\epsscale{1.15}
\plotone{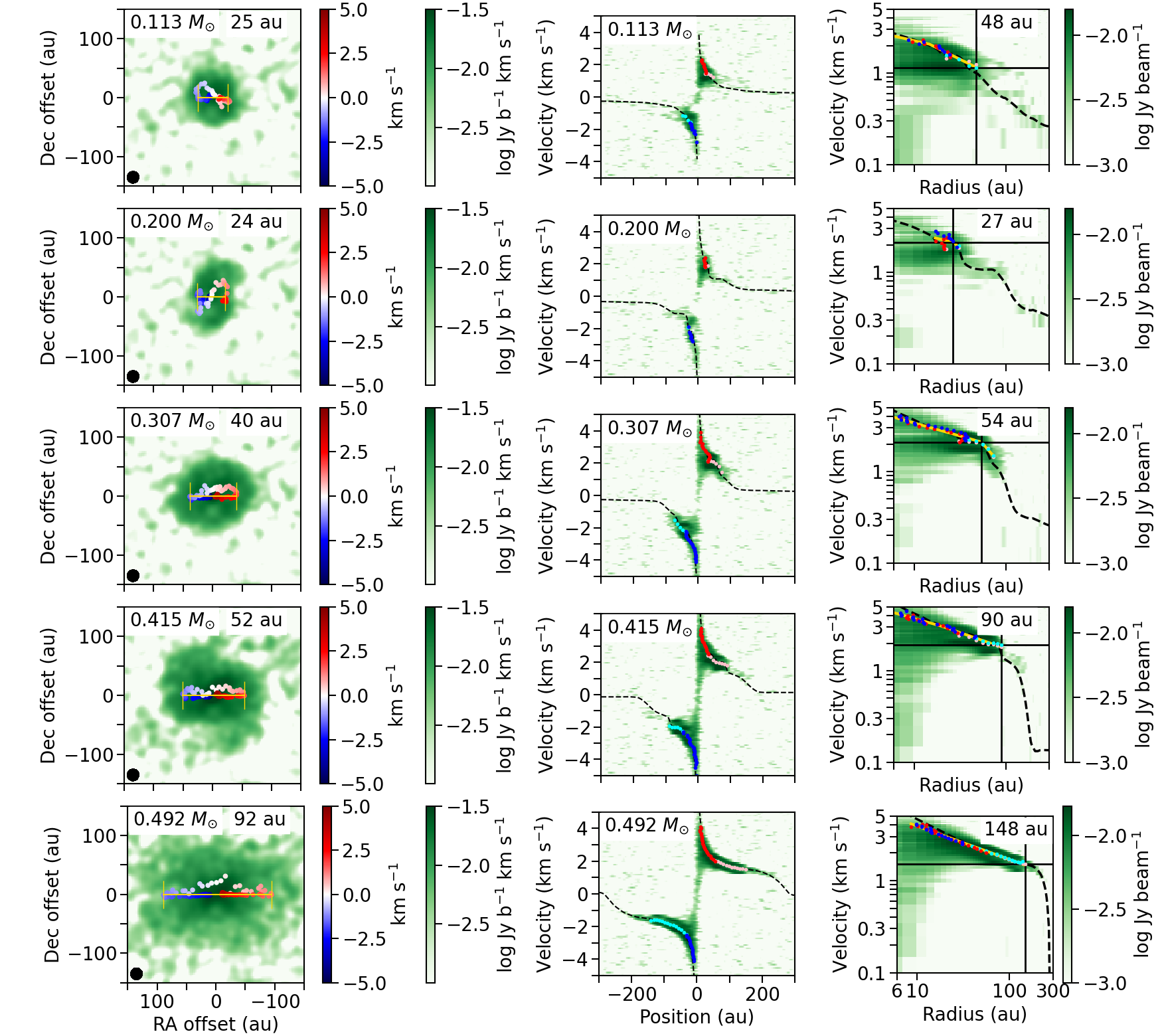}
\caption{
Same as the right panels of Figures~\ref{fig:chan} and \ref{fig:pv} but with a more realistic treatment of noise and continuum subtraction, in which the continuum visibilities are subtracted after artificial noise is added to the visibilities and then the data set is CLEANed.
\label{fig:reall}
}
\end{figure}

Figures~\ref{fig:realc} and \ref{fig:reall} plot the results of the five representative epochs in the same format as in \S\ref{sec:ana}. 
The overall features seen in these figures are almost the same as in their counterparts in \S\ref{sec:ana}.
The measured radii plotted in these figures agree well  with those derived in \S\ref{sec:ana} within the uncertainties due to the noise. 
For example, the radii estimated from the continuum emission in \S\ref{sec:ana} are $R_{\rm img}=13$, 16, 24, 34 and 61\,au at the epochs of 0.113, 0.200, 0.307, 0.415, and $0.492~\Ms$.
On the other hand, the radii derived from the realistic noise inclusion and continuum subtraction are $R_{\rm img}=14$, 16, 24, 34 and 58\,au at the same epochs (Fig.~\ref{fig:realc}).
Thus, the difference between them is not significant.  

\end{document}